\documentclass{elsart}
\usepackage[english]{babel}
\usepackage{graphicx,subfig,placeins}
\usepackage{amsfonts}
\usepackage{amsmath}
\usepackage{color}
\usepackage{hyperref}

\usepackage{multirow} 

\newcommand{\diag}{\text{diag}}
\newcommand{\dep}{\text{dep}}
\newcommand{\dist}{\text{dist}}

\input{tcilatex}

\begin{document}

\begin{frontmatter}%

\title{Spectra and pseudospectra in the evaluation of material stability in phase field schemes}%

\author{Michele Benzi}%
\address{Scuola Normale Superiore\\Piazza dei Cavalieri 7, I-56126 Pisa, Italy\\e-mail: michele.benzi@sns.it}%

\author{Daniele La Pegna}%
\address{Scuola Normale Superiore\\Piazza dei Cavalieri 7, I-56126 Pisa, Italy\\e-mail: daniele.lapegna@sns.it}%

\author{Paolo Maria Mariano}
%

\address{DICEA, Universit\`{a} di Firenze\\via Santa Marta 3, I-50139 Firenze, Italy\\e-mail: paolomaria.mariano@unifi.it}%





\begin{abstract}
We consider the dynamics of bodies with ‘active’ microstructure described by vector-valued phase
fields. For waves with time-varying amplitude, the associated evolution equation involves a matrix
that can be non-normal, depending on the constitutive choices adopted for the microstructural
actions associated with the considered phase field.
The occurrence of non-normality requires to look at the pseudospectrum of the considered matrix, namely the
set of all possible eigenvalues of matrices in a $\varepsilon$-neighborhood of the matrix itself, because the eigenvalues of non-normal matrices can be very sensitive to small perturbations and therefore the
spectral analysis alone would not be sufficient to distinguish with certainty between stable and unstable behavior. We develop the relevant analyses in the case of quasicrystals for which the values of some constitutive parameters are not known or are uncertain from an experimental point of view, a circumstance suggesting parametric analyses. We find circumstances in which the pseudospectra obtained by means of the so-called structured perturbations predict instability when, instead, the spectral analysis indicates stability.
\end{abstract}%


\end{frontmatter}%

\section{Introduction}

We discuss the propagation of waves with variable amplitude in bodies with \textquoteleft active' microstructure described by vector-valued phase fields.
The adjective \textquoteleft active' refers to bodies with microstructure endowed with inner (but observable) degrees of freedom with pertinent interactions not directly associated with those kinematic mechanisms that are commonly associated with the standard stress (namely, the crowding and shearing of material elements merely considered as black boxes).
In particular, we investigate the influence of a microstructural self-action on the stability of the material. Non-normal matrices emerge along the analysis. Their occurrence requires supplementing the conventional spectral approach to stability with an analysis of pseudospectra of the operators involved to verify stability and possible transitions to unstable behavior with certainty.
Indeed, there may be conditions (expecially when constitutive parameters are uncertain or time-varying) in which pseudospectra indicate instability while, on the contrary, the standard spectral analysis predicts stability. In order to show concretely this phenomenon, in this paper we specifically refer to quasicrystals, but the approach has general validity.

Quasicrystals are materials with perfect long-range order, but with no three-dimensional translation periodicity \cite{Shecht84}. Their structure is made of atomic clusters distributed with a symmetry that is incompatible with a periodic tiling of the ambient space, for example icosahedral symmetry in $3D$-space \cite{Shecht84}, \cite{RWM88}. The quasi-periodic lattice is characterized by additional low-spatial-scale degrees of freedom, so-called \emph{phasons}, here indicated by $\nu$ at $x$ and $t$.
They are justified on geometrical basis. Indeed, if we expand in Fourier series the mass distribution of a $3D$ quasi-periodic lattice, the resulting wave vectors are six-dimensional \cite{RWM88}. More generally, a quasi-periodic lattice in a $n$-dimensional point space can be seen as a projection of a periodic lattice in a higher dimensional ambient onto a $nD$ incommensurate subspace. To visualize the circumstance, consider a $2D$ lattice with square symmetry; take in the plane a straight line inclined by an irrational angle with the lattice symmetry axes and project over it, orthogonally, the lattice atoms in a strip around the line (whatever be its thickness); the result is a quasi-periodic $1D$ lattice over the line, which is in this case the incommensurate subspace.
Thus, a displacement in the higher-dimensional periodic lattice has a component in the incommensurate subspace and one orthogonal to it; the first projection of the displacement field is what we commonly call a \emph{phonon field}, because it is associated with acoustic waves; the orthogonal displacement component to the incommensurate subspace leads to the field of \emph{phasons}. This geometrical aspect first suggested a convenient description of the quasicrystal behavior under external actions as a higher-dimensional replica of the mechanics of simple bodies \cite{FAN}, \cite{HU}.
However, if we consider the degrees of freedom represented by $\nu$ as those exploited by microscopic rearrangements that assure quasi-periodicity of the atomic lattice under given (possibly time-varying) boundary conditions, at continuum scale they are inner to every material element associated with a geometric point; so, $\nu$ itself has to be considered insensitive to relative rigid translations of observers in the physical space while, being a vector, it is sensitive to observer rotations. On the other side, even if we do not accept such an interpretation, since phasons belong to the space orthogonal to the incommensurate subspace (the physical space) on which we project the hyperlattice (that is the lattice in the higher-dimensional space), for a mere geometrical reason phasons are insensitive to observer translations in the physical space. All this implies that the mechanics of quasicrystals is more than a replica of the mechanics of simple bodies rewritten in a higher dimensional space.
Rather it falls within the general model-building framework for the mechanics of complex bodies (as shown in \cite{M06}, \cite{MP13}).
In this general setting, balance equations are those of forces and those of microstructural interactions associated with the phase field; a balance of couples links standard and microstructural stresses, and a self-action at microstructure level. 
Although these balance equations can be postulated as first principles \cite{C89}, they can be derived from invariance principles together with the representation of contact actions in terms of standard and microstructural stresses.
If we consider even the fully dissipative setting,
the principles mentioned are the requirement of objectivity for the external power alone \cite{M02} or the covariance (that is, structure-invariance under diffeomorphism-based observer changes) of the Clausius-Duhem inequality, written in terms of external power \cite{M24}.
The first path does not include the case in which the phase field is scalar or pseudo-scalar, the latter path covers even such cases.
Resorting to first principles allows an appropriate control on the results unlike a mere use of analogies that can be surely useful but, in essence, are only hopes, although at times concrete ones.

As regards quasicrystals, the invariance techniques mentioned (and here we adopt the $SO(3)$ invariance of the external power alone) imply the possible emergence of a phason self-action $\mathbf{z}$ which can have both conservative (energetic-type) and dissipative components \cite{MP13}, the latter indicated as phason friction \cite{RL02}. The conservative component of $\mathbf{z}$ plays a role in the stability of the material.
To explore the circumstance, we analyze the time-dependent amplitude wave propagation in different cases:
\begin{enumerate}
\item[1)] absence of a self-action, namely $\mathbf{z}=0$;
\vskip6pt
\item[2)] purely conservative character of the self-action, meaning that $\mathbf{z}$ equals the derivative of a free energy density $\psi$ with respect to $\nu$; a special choice is $\mathbf{z}=k_0 \nu$, $k_0\in\mathbb{R}^{+}$;
\vskip6pt
\item[3)] purely dissipative self-action, so that $\mathbf{z}$ represents microstructural friction alone; thus, it is such that $\mathbf{z}\cdot \dot{\nu} \geq 0$, for any $\dot{\nu}$, an inequality that is compatible with the choice $\mathbf{z} = \varsigma\dot{\nu}$, $\varsigma\in\mathbb{R}^{+}$;
\vskip6pt
\item[4)] simultaneous presence of both conservative ($\mathbf{z}^e$) and dissipative ($\mathbf{z}^d$) components of the self-action, namely $\mathbf{z}=\mathbf{z}^e + \mathbf{z}^d$, where $\mathbf{z}^e = k_0 \nu$ and $\mathbf{z}^d = \varsigma\dot{\nu}$.
\end{enumerate}
By spectral and pseudospectral analyses of the system that governs the wave amplitude evolution, we determine for the experimentally unknown constitutive parameters possible thresholds for stable-unstable transition in the material behavior.

The notion of pseudospectrum emerged from analyses in fluid dynamics and was later adopted in various other fields, from laser physics, control theory and others.
Such a method deals with the spectral analysis of non-normal matrices. Precisely, in general, with $A\in\mathbb{C}^{n\times n}$ a $n\times n$ matrix over the field of complex numbers, and $A^{\ast}$ its conjugate transpose, we say that $A$ is non-normal when $AA^{\ast}\neq A^{\ast}A$. We can also evaluate the degree of non-normality \cite{Els87}. However, besides formal definitions, let us come back to physics.

To evaluate the stability of fluid flows, we commonly (\emph{i})  linearize the balance equations about the laminar state and (\emph{ii}) look at the spectrum of the linearized operator to check whether there are unstable eigenvalues (those with real part in the positive half-plane). The method applies safely to various cases of simple fluids such as B\'{e}nard convection and Taylor-Couette flow. However, for the plane Couette flow the approach predicts linear stability for all Reynolds numbers, but experiments show transition to turbulence at Reynolds numbers as low as 350. Another failure in the theoretical stability predictions occurs, for example, in the analysis of the plane Poiseuille flow. Indeed, although the eigenvalues of the linearized operator lie in the stable half-plane, transient energy grows \cite{TTRD93}. Non-orthogonality of eigenstates, which is a manifestation of non-normality of the governing operators,
occurs in nuclear scattering, electronic conductance, and wave propagation in disordered media \cite{DaG19}. Also, convection-diffusion problems (damage evolution is an example) can be highly non-normal as many of the problems involving mixed derivatives \cite{ReT94}.

In these cases, reliable stability information can be obtained by consideration of the $\varepsilon$-pseudospectrum (see, e.g., \cite{TT87}, \cite{TE}, \cite{GvdV00}). A complex number $z\in\mathbb{C}$ is in the $\varepsilon$-pseudospectrum of a matrix or, more generally, an operator $A$ if it is in the spectrum of some perturbed operator $A+B$ with $||B||<\varepsilon$. Pseudospectral material techniques for fluids have been largely developed for the Navier-Stokes equations (see, e.g., \cite{GRW12}, \cite{KHT87}, \cite{Pri95}) and a few other problems \cite{TE}).

The spectral analysis of the acoustic tensor is the key ingredient for evaluating the material stability in solids. Such a tensor is symmetric in the traditional setting of continuum mechanics; the matrix of its components has only real entries. This is essentially due to the symmetry of the Cauchy stress. The property obviously remains when we homogenize simple lattices, which lead to a traditional continuum scheme. In both cases, the spectral analysis generically suffices to evaluate the material stability.

When we consider bodies with active microstructure, as already mentioned, phase fields enter into the representation of the body morphology and are intended as observable entities. Interactions are associated with these fields and satisfy appropriate balance equations. Accounting for these interactions generically implies a lack of symmetry for the Cauchy stress \cite{C89}, \cite{M02}. We also obtain this scheme when we homogenize complex lattices in which we have internal degrees of freedom (think for example of a lattice in which every material point is a small rigid body free to rotate about its mass center; the homogenized continuum is a Cosserat's one), so that the homogenized continuum is not a classical (Cauchy-type) one but involves phase fields describing the additional degrees of freedom (rotations in the Cosserat case) \cite{M22}. However, in the linearized setting, even in the presence of microstructural actions the Cauchy stress reacquires always the symmetry \cite{M24}. This aspect notwithstanding, admissible constitutive relations for the microstructural interactions may induce non-normality, as we show in this paper.

Thus, when non-normality occurs, due to microstructural interactions, a pseudospectral analysis allows a refined (and reliable) evaluation of the material stability.
To the best of our knowledge, we adopt for the first time the pseudospectral analysis in the mechanics of solids. Parametric analyses due to uncertainty and lack of knowledge of the constitutive parameters show circumstancxes in which the psedudospectra indicate possible instability while, instead, the spectral analysis show stability. We also highlight the important role played by (real) structured perturbations in addition to the more commonly used generic (complex) perturbations.

The paper is organized as follows: in \textbf{Section 2} we indicate the basic fields associated with the body morphology. In \textbf{Section 3} we discuss the notion of observer and obtain the balance equations, including those of the microstructural interactions. We avoid merely writing them down as postulates; we \emph{derive} them from a requirement of objectivity of the external power to show clearly the origin and type of the lack of symmetry of the Cauchy stress. In \textbf{Section 4} we discuss constitutive structures. In \textbf{Section 5} we analyze the propagation of waves with variable amplitude in the linearized setting. To prepare and clarify the pseudospectral analysis, in \textbf{Section 6} we recall basic notions and properties of the $\varepsilon$-pseudospectrum. Numerical results are discussed in \textbf{Section 7}. Some coefficients used in numerical simulations are known from experimental data. For those that are experimentally unknown, we perform parametric analyses in the admissible ranges that are theoretically established.

We refer all the following analyses to orthonormal frames of reference for the sake of simplicity. So, we do not distinguish between covariant and contravariant components of tensors; similarly we avoid distinguishing between transpose (indicated by a superscript $\top$) and formal adjoint (commonly denoted by an asterisk alone) of linear operators. 

An interposed dot will indicate the scalar product between elements of a linear space, whatever it be.
As a further matter of notation, for a generic vector $k\in\mathbb{R}^{n}$, the symbol $|k|$ will indicate as usual the modulus of $k$. For a generic matrix $A$ with entries $a_{ij}$, $\Vert A\Vert$ will indicate the value of a generic norm; $\Vert \bullet\Vert_2$ will be written for the 2-norm, so that $\Vert A\Vert_2=s_{max}(A)$, where $s_{max}(A)$ is the maximum singular value of $A$; the symbol $\Vert \bullet\Vert_{F}$ will be used for the Frobenius norm defined by $\Vert A\Vert_{F}=\sqrt{\mathrm{tr}(A^{\ast}A)}$, where $\mathrm{tr}$ is the trace operator. The summation over repeated indices is occasionally adopted throughout the paper.

\section{Basic fields} \label{Se2}

Consider two isomorphic copies of the $3D$ real space, namely $\hat{\mathbb{R}}^{3}$ and $\mathbb{R}^{3}$;
the isomorphism 
$\hat{\iota}:\hat{\mathbb{R}}^{3}\longrightarrow\mathbb{R}^{3}$ is merely the identification map.
In $\hat{\mathbb{R}}^{3}$ we select a fit region $\mathfrak{B}$, a bounded simply connected domain endowed with a surface-like boundary oriented by the outward unit normal $\mathbf{n}$ to within a finite number of corners and edges. We adopt $\mathfrak{B}$ as a macroscopic reference
configuration.
From it, deformations are orientation-preserving differentiable one-to-one maps
$x\longmapsto y:=\tilde{y}(x)\in\mathbb{R}^{3}$.  We indicate by $F$ the deformation gradient.
To account for motions in a time interval $[0,\bar{t}\,]$, we extend $\tilde{y}$ to the space-time tube $\mathfrak{B}\times[0,\bar{t}\,]$ so that we have $(x,t)\longmapsto y:=\tilde{y}(x,t)\in\mathbb{R}^{3}$ and we take $\tilde{y}(x,\cdot)$ to be differentiable also with respect to time at every $x\in\mathfrak{B}$; a superposed dot will indicate the time derivative. We thus define a \emph{displacement} $\tilde{u}$ in the physical space as
\[
(x,t)\longmapsto u:=\tilde{u}(x,t)=\tilde{y}(x,t)-\hat{\iota}(x)\;.
\]
Thus, we have $F=I+\nabla u$, where $I$ is the shifter from $\hat{\mathbb{R}}^{3}$ to $\mathbb{R}^{3}$,
with components $\delta^{i}_{\;A}$, where capital indices refer to
coordinates in $\hat{\mathbb{R}}^{3}$, while the others to their
counterparts in $\mathbb{R}^{3}$.

In Lagrangian representation the velocity is $\dot{y}=\frac{\partial\tilde{y}(x,t)}{\partial t}=\dot{u}$.

A differentiable phase field
\[
(x,t)\longmapsto \nu:=\tilde{\nu}(x,t)\in\tilde{\mathbb{R}}^{3}
\]
brings at macroscopic scale information on microstructural events at spatial scale $\lambda$; its choice is matter of modeling, so it depends on the cases considered. In the special case of quasicrystals considered here, the ratio between the modulus of $\nu$ and the one of $u$ is about $10^{-2}$, as shown by numerical simulations in \cite{MS15} on the basis of experimental data.

$\tilde{\mathbb{R}}^{3}$ is isomorphic to $\mathbb{R}^{3}$ but distinguished from it; the isomorphism is once again the identification.
We indicate by $N$ its spatial derivative, namely
\[
N:=\nabla\nu=\bigg(\frac{\partial\nu^{i}}{\partial x}\bigg)^{A}\tilde{\mathbf{e}}_{i}\otimes\mathbf{e}_{A}\,,
\]
where $\tilde{\mathbf{e}}_{i}$ is the $ith$ element of a dual basis over $\tilde{\mathbb{R}}^{3}$ and $\mathbf{e}^{A}$ is the $Ath$ element of the dual basis over $\hat{\mathbb{R}}^{3}$. The time rate of $\nu$ in Lagrangian representation is given by
$\dot{\nu}=\frac{\mathrm{d}\tilde{\nu}(x,t)}{\mathrm{d}t}$.

The small strain regime is commonly defined by the condition $\Vert\nabla u\Vert_F\ll1$. In such a regime we do not distinguish between $\hat{\mathbb{R}}^{3}$ and $\mathbb{R}^{3}$.

\section{Invariance and balance}

\subsection{Observers and their changes}

The prescription of reference frames over \emph{all} the spaces used for representing the morphology of a body and its motion is what we consider to be an \emph{observer}.
Here such spaces are the reference one (namely $\hat{\mathbb{R}}^{3}$), the physical space $\mathbb{R}^{3}$, the one where $\tilde{\nu}$ takes values, namely $\tilde{\mathbb{R}}^{3}$, and the time interval.
We adopt observer changes that leave invariant the time scale and $\hat{\mathbb{R}^{3}}$, while they are isometries in $\mathbb{R}^{3}$, with consequent changes of reference frames on $\tilde{\mathbb{R}}^{3}$.

More precisely, if $y$ is a place evaluated by a first observer $\mathcal{O}$, another observer $\mathcal{O}^{\prime}$ records a counterpart of $y$ given by $y'=a(t)+Q(t)(y-y_{0})+y_{0}$, where $t\longmapsto a(t)\in\mathbb{R}^{3}$ is a time-dependent vector-valued smooth map; $y_{0}$ is an arbitrary fixed point, and $t\longmapsto Q(t)\in SO(3)$ is a time-dependent orthogonal tensor-valued smooth map.

$\mathcal{O}$ records a velocity $\dot{y}$, which is $\dot{y}'=\dot{a}+\dot{Q}(y-y_{0})+Q\dot{y}$ for $\mathcal{O}^{\prime}$. By pulling back $\dot{y}'$ to $\mathcal{O}$, we get $\dot{y}^{\diamond}:=Q^{\mathsf{T}}\dot{y}'$ given by
$$\dot{y}^{\diamond}=c+q\times(y-y_{0})+\dot{y}\;.$$
The vector $c:=Q^{\mathsf{T}}\dot{a}$ is a relative translation velocity between the two observers, while $q$ is the axial vector of the skew-symmetric second-rank tensor $Q^{\mathsf{T}}\dot{Q}$.

The vector $\nu:=\tilde{\nu}(x,t)\in\tilde{\mathbb{R}^{3}}$ collects degrees of freedom that are \emph{internal} (thus \emph{relative}) in time to the material element associated with $x\in\mathfrak{B}$ in the reference configuration. Thus $\nu$ is a Galilean invariant vector, meaning it is insensitive to relative rigid translations between two generic observers in the physical space.
When $\mathcal{O}$ and $\mathcal{O}^{\prime}$ rotate relatively, the value $\nu$ recorded by a first observer changes into $\nu'=Q(t)\nu$ and the time rate
$\dot{\nu}$ becomes $\dot{\nu}'=\dot{\nu}+\dot{Q}\nu$.
By pulling back $\dot{\nu}'$ into $\mathcal{O}$, we obtain a new vector $\dot{\nu}^{\diamond}:=Q^{\mathsf{T}}\dot{\nu}'$ given by
$$\dot{\nu}^{\diamond}=\dot{\nu}+q\times\nu\;,$$
which we rewrite as
$$\dot{\nu}^{\diamond}=\dot{\nu}+\mathcal{A}(\nu)q\;,$$
where the linear operator $\mathcal{A}$ is given by
$\mathcal{A}:=\mathcal{A}(\nu)=-\nu\times$.

\subsection{External power and invariance requirement: consequences}

A subset $\mathfrak{b}$ of $\mathfrak{B}$ is said to be a \emph{part} of
$\mathfrak{B}$ when it has non-vanishing volume and the same geometrical regularity required of $\mathfrak{B}$.

External actions over $\mathfrak{b}$ are associated with $u$ and $\nu$; they are defined by the power that they perform in the time rates of the two fields considered. Such actions are subdivided, as usual, into bulk and contact families: those associated with $u$ are the standard bulk force and boundary traction, while those performing power in $\dot{\nu}$ are the microstructural actions. 
The \emph{external power}
$\mathcal{P}^{ext}_{\mathfrak{b}}$ that they perform on any time rate of
change in the body morphology is defined in Lagrangian representation to be
\begin{equation*}
\mathcal{P}_{\mathfrak{b}}^{ext}\left( \dot{y},\dot{\nu}\right) :=\int_{%
\mathfrak{b}}\left( b^{\ddagger }\cdot \dot{y}+\beta ^{\ddagger }\cdot \dot{%
\nu}\right) \mathrm{d}x+\int_{\partial \mathfrak{b}}\left( \mathfrak{t}_{\partial}\cdot \dot{y}%
+\tau_{\partial} \cdot \dot{\nu}\right) \mathrm{d}\mathcal{H}^{2}(x),
\end{equation*}
where
$\mathrm{d}\mathcal{H}^{2}(x)$ denotes the Hausdorff two-dimensional measure. The symbol $\partial$ as a subscript indicates that the contact actions defined over $\partial \mathfrak{b}$ depend on the boundary, in addition to $x$ and $t$. The bulk force $b^{\ddagger }$ is presumed to be the sum of inertial $b^{in}$ and non-inertial $b$ components, the former to be identified; $\beta ^{\ddagger }$ also admits by assumption a similar decomposition: $\beta ^{\ddagger }=\beta^{in}+\beta$.

We require that $\mathcal{P}_{\mathfrak{b}}^{ext}$ is invariant under isometry-based observer changes; so, we impose
$$\mathcal{P}_{\mathfrak{b}}^{ext}\left( \dot{y},\dot{\nu}\right)
=\mathcal{P}%
_{\mathfrak{b}}^{ext}\left( \dot{y}^{\diamond },\dot{\nu}^{\diamond
}\right)$$
for any choice of $\mathfrak{b}$,
$c$,\emph{\ and }$q$. It means that we consider $\mathcal{P}_{\mathfrak{b}}^{ext}$ to be objective.
The arbitrariness of $c$ implies the standard integral balance of forces
\begin{equation}
\int_{\mathfrak{b}}b^{\ddag }\;\;\mathrm{d}x+\int_{\partial \mathfrak{b}}\mathfrak{t}_{\partial}\;\;\mathrm{d}
\mathcal{H}^{2}(x)=0\;,  \label{IBF}
\end{equation}
while the one of $q$ a non-standard balance of couples
\begin{equation}
\int_{\mathfrak{b}}\left( \left( y-y_{0}\right) \times b^{\ddag }+\mathcal{A}%
^{\top }\beta ^{\ddag }\right) \mathrm{d}x+\int_{\partial \mathfrak{b}}\left( \left(
y-y_{0}\right) \times \mathfrak{t}_{\partial}+\mathcal{A}^{\top }\tau_{\partial} \right) \mathrm{d}\mathcal{H}^{2}(x)=0\;.  \label{IBC}
\end{equation}

Equation \eqref{IBF} gives standard results. Precisely, if $| b^{\ddag }| $ is bounded over $\mathfrak{B}$
and $\mathfrak{t}_{\partial}$ depends continuously on $x$, Cauchy's theorem on fluxes implies that  $\mathfrak{t}_{\partial}$ depends
on $\mathfrak{\partial\mathfrak{b}}$ only through the normal $\mathbf{n}$ to it in all points where $\mathbf{n}$ is well-defined; $\mathfrak{t}_{\partial}$ is also such that
$\mathfrak{t}_{\partial}=\mathfrak{t}:=\tilde{\mathfrak{t}}\left( x,t,\mathbf{n}\right) =-\tilde{\mathfrak{t}}\left( x,t, -\mathbf{n}\right)$, and, as a function of $\mathbf{n}$, $\tilde{\mathfrak{t}}$ is homogeneous and
additive, namely $\tilde{\mathfrak{t}}\left( x,t,\mathbf{n}\right) =P\left( x,t,\right) \mathbf{n}\left( x\right)$, with $P$ the first Piola-Kirchhoff stress.

Less standard are the consequences of the integral balance \eqref{IBC}. Precisely,
since $\mathfrak{B}$ is bounded, we can choose the arbitrary point $y_{0}$ in a way such that the boundedness of $| b^{\ddag }| $ implies the one of $| (y-y_{0})\times b^{\ddag }|$. If $| \mathcal{A}^{\top }\beta ^{\ddag }| $
is also bounded and $\tau_{\partial}$ depends continuously on $x$, by exploiting once again Cauchy's theorem, we realize that $\tau_{\partial}$ depends on $\mathfrak{\partial\mathfrak{b}}$ only through the normal $\mathbf{n}$; also,
$\mathcal{A}^{\top }\left( \tilde{\tau} \left( x,t,\mathbf{n}\right) +\tilde{\tau} \left( x,t,-\mathbf{n}\right)
\right) =0$ and, as a function of $\mathbf{n}$, $\tilde{\tau}$ is homogeneous and additive:  $\tilde{\tau} \left( x,t,\mathbf{n}\right) =\mathcal{S}\left( x,t\right) \mathbf{n}\left( x\right)$. The second-rank tensor $\mathcal{S}$ is named a \emph{microstress}.

Due to the arbitrariness of $\mathfrak{b}$, if both stress fields are in $C^{1}\left( \mathfrak{B}\right) \cap C\left( \bar{\mathfrak{B}}\right) $ and
the bulk actions $x\longmapsto b$, $x\longmapsto \beta ^{\ddag }$\ are continuous
over $\mathfrak{B}$, from \eqref{IBF} we get the standard point-wise balance of forces
\begin{equation}
\mathrm{Div}P+b^{\ddag }=0\;,  \label{pointCau}
\end{equation}
while from \eqref{IBC} we get that there exists a field $(x,t)\longmapsto \mathbf{z}\left( x,t\right) \in
\hat{\mathbb{R}}^{3}$ such that
\begin{equation}
\mathrm{Div}\mathcal{S}+\beta ^{\ddag }-\mathbf{z}=0\;,\qquad \text{and}\qquad\mathrm{skw}(PF^{\top })=\frac{1}{2}\mathsf{e}\left( \mathcal{A}^{\top
}\mathbf{z}+\left( \nabla\mathcal{A}^{\top }\right) \mathcal{S}\right)\;, \label{f3}
\end{equation}
where $\mathrm{skw}(\cdot)$ extracts the skew-symmetric part of its argument; moreover,
\begin{equation}
\mathcal{P}_{\mathfrak{b}}^{ext}\left( \dot{y},\dot{\nu}\right) =\int_{%
\mathfrak{b}}\left( P\cdot \dot{F}+\mathbf{z}\cdot \dot{\nu}+\mathcal{S}\cdot \dot{N}%
\right) \mathrm{d}x\;,  \label{Inner}
\end{equation}
where the right-hand side integral is called \emph{internal }(or \emph{inner}) \emph{power} (see details in the case of general manifold-valued phase fields in \cite{M02} and \cite{M16}).

The inertial terms are identified by assuming that their power equals the negative of the kinetic energy time rate for any choice of the velocity fields. This approach is common. The kinetic energy of a generic complex body is presumed to be the sum of a standard quadratic expression $\frac{1}{2}\rho|\dot{y}|^{2}$, where $\rho$ is the mass density, plus a non-negative function $\kappa(\nu, \dot{\nu})$ such that $\kappa(\nu, 0)=0$ and $\frac{\partial^{2}\kappa}{\partial\dot\nu\partial\dot\nu}\cdot\dot\nu\otimes\dot\nu\geq 0$ for generic phase fields \cite{C89} (for more general analyses on $\beta^{\ddag }$ see \cite{CGio}, \cite{M02}, \cite{M16}).

In the case of quasicrystals that we treat here with details, $\tilde{\nu}$ coincides with the so-called \emph{phason field}; for such a material the possibility of peculiar inertia effects attributed to $\tilde{\nu}$ is questionable since only three sound-like branches are apparently recorded in experiments \cite{SvS}. Consequently, even if only for later purposes, we leave apart the microstructural inertia and set $\kappa(\nu, \dot{\nu})=0$, so $\beta^{in}=0$ to within a powerless term that we also discard (for its presence in quasicrystals see \cite{MP13} and \cite{BM22}). The gross kinetic energy remains; the arbitrariness of $\dot{y}$ implies by continuity the identification  $b^{in}=-\rho \ddot{y}=-\rho\ddot{u}$
to within a powerless term that we also discard assuming that we are using an inertial frame of reference (in turn we can say that a frame of reference is inertial when the above identification of $b^{in}$ is exact).

\section{Constitutive structures} \label{Se4}

We write the Clausius-Duhem inequality in isothermal setting as
\begin{equation*}
\frac{\mathrm{d}}{\mathrm{d}t}\int_{\mathfrak{b}}\psi\;\;\mathrm{d}x-\int_{%
\mathfrak{b}}\left( P\cdot \dot{F}+\mathbf{z}\cdot \dot{\nu}+\mathcal{S}\cdot \dot{N}%
\right) \mathrm{d}x\leq 0\;,
\end{equation*}
where $\psi$ is the \emph{free energy density}.
The inequality, a version of the second law of thermodynamics, is presumed to hold true for any choice of the time rates involved.

The arbitrariness of $\mathfrak{b}$ and the presumed continuity of the integrand imply a local dissipation inequality given by
$$\dot{\psi}-P\cdot\dot{F}-\mathbf{z}\cdot\dot{\nu}-\mathcal{S}\cdot\dot{N}\leq 0\;.$$

As constitutive functional dependence on
state variables, we assume
$$\psi=\tilde{\psi}(F,\nu, N)\;\qquad P=\tilde{P}(F,\nu, N)\;\qquad \mathcal{S}=\tilde{\mathcal{S}}(F,\nu, N)$$
and
$$\mathbf{z}=\tilde{\mathbf{z}}^{e}(F,\nu, N)+\tilde{\mathbf{z}}^{d}(F,\nu, N, \dot{\nu})\;,$$
where, we recall, $\mathbf{z}^{d}=\tilde{\mathbf{z}}^{d}(F,\nu, N, \dot{\nu})$ is a dissipative component of the self-action $\mathbf{z}$, while $\mathbf{z}^{e}=\tilde{\mathbf{z}}^{e}(F,\nu, N)$ is a conservative component determined by the free energy $\psi$.

These choices imply that we consider dissipation to be restricted only at microstructural level. Thus, arbitrariness of the time rates involved in the inequality necessarily implies
\begin{equation*}
P=\frac{\partial\psi}{\partial F}\;,\quad\mathbf{z}^{e}=\frac{\partial\psi}{\partial\nu}\;,
\quad\mathcal{S}=\frac{\partial\psi}{\partial N}\;,\quad\mathbf{z}^{d}\cdot\dot{\nu}\geq 0.
\end{equation*}
The last inequality is compatible with
\begin{equation*}
\mathbf{z}=\frac{\partial\psi}{\partial\nu}+\tilde{\varsigma}(\dots)\dot{\nu}\;,
\end{equation*}
where $\tilde{\varsigma}(\dots)$ is a
positive-valued state function, here chosen to be a scalar
$\varsigma>0$ only for the sake of simplicity. With $\nu$ referring to a length scale $\lambda$ (as already recalled in Section \ref{Se2}), both $\mathcal{S}$ and $\mathbf{z}$ are thus of order $\lambda$. Consequently, from equation \eqref{f3}, we obtain that $PF^{\top}$ is symmetric to within terms of the order $\lambda^{2}$.

$P$, $\mathcal{S}$, and $\mathbf{z}$ are values of fields defined over $\mathfrak{B}$ in terms of space variables. They have counterparts defined over 
$\mathfrak{B}_{c}:=\tilde{y}(\mathfrak{B},t)$ and given by
\begin{equation*}
\boldsymbol{\sigma}=\frac{1}{\mathrm{det} F}\frac{\partial\psi}{\partial F}F^{\top}\;,\quad \mathbf{z}_{c}
=\frac{1}{\mathrm{det} F} \left(\frac{\partial\psi}{\partial\nu}+\tilde{\varsigma}(\dots)\dot{\nu}\right) \;,\quad
\mathcal{S}_{c}=\frac{1}{\mathrm{det} F}\frac{\partial\psi}{\partial N}F^{\top},
\end{equation*}
where $\sigma$ is the standard \emph{Cauchy stress}, while $\mathbf{z}_{c}$ and $\mathcal{S}_{c}$ are respectively self-action and microstress in Eulerian representation.

In small strain regime, that is, we recall, when $\|\nabla u\|_F\ll 1$, we have the approximations
\begin{equation*}
\boldsymbol{\sigma}\approx P\;,\quad \mathbf{z}_{c}\approx \mathbf{z}\;,\quad\mathcal{S}_{c}\approx \mathcal{S}\;.
\end{equation*}
It is also possible to choose for the free energy
density $\psi$ a quadratic form with respect to its entries, a choice otherwise forbidden at least with respect to $F$ because, as it is well-known, a quadratic dependence on $F$ would be incompatible with the natural request
of objectivity for the energy density, that is a requirement
of invariance under the action of $SO(3)$, hence with respect to relatively rotating observers.

By maintaining ourselves in the small strain setting, we thus choose a form of energy derived in \cite{MP13} under symmetry requirements. It reads
\begin{equation*}
\begin{aligned}
\psi =&\frac{1}{2}\lambda \left( \mathrm{sym} \nabla u \cdot
I\right)^{2}
+\mu\, \mathrm{sym} \nabla u \cdot \mathrm{sym}\nabla u  \\
&+\frac{1}{2}k_{1}\left( N\cdot I\right)^{2}
+k_{2}\,\mathrm{sym}N\cdot
\mathrm{sym}N+k_{2}^{\prime }\,\mathrm{skw}N\cdot \mathrm{skw}N \\
&+k_{3}\left( \mathrm{sym}\nabla u \cdot I\right) \left( N\cdot I\right)
+k_{3}^{\prime }\,\mathrm{sym}N\cdot \mathrm{sym}\nabla u
+\frac{1}{2}k_{0}\left\| \nu \right\| ^{2}
\end{aligned}
\end{equation*}
where $I$ is once again the unit tensor and  $\mathrm{sym}(\cdot)$
extracts the symmetric component of its argument; $\lambda$ and $\mu$ are the
Lam\'{e} constants, the other coefficients indicated by $k$ with various decorations are related to $\nu$. They
must be such that the energy is non-negative definite, thus
$\mu > 0$, $k_2 > 0$, $2 k_2 + 3k_1 > 0$, $k_2' > 0$, $k_3'  < 2\sqrt{\mu k_2}$, $3 k_3 + k_3' < \sqrt{(2\mu+3\lambda)(2k_2+3k_1)}$, $k_{0}\geq 0$.

We thus have
\begin{equation*}
\begin{aligned}
& P\approx\boldsymbol{\sigma} =\lambda \left( \mathrm{tr}\, (\mathrm{sym}\nabla u) \right)
I+2\mu \mathrm{sym}\nabla u
+k_{3}\left( \mathrm{tr}\nabla\nu\right) I+k_{3}^{\prime }\mathrm{sym}\nabla\nu\;,\\[0.1 cm]
&\mathbf{z}\approx \mathbf{z}_{c}=k_{0}\nu+\varsigma\dot{\nu}\;,\\[0.1 cm]
& \!\!\begin{array}{ll}\mathcal{S}\approx\mathcal{S}_{c}= &k_{1}\left( \mathrm{tr}\nabla\nu\right)
I+2k_{2}\mathrm{sym}%
\nabla\nu
+2k_{2}^{\prime }\mathrm{skw}\nabla\nu+k_{3}\left(
\mathrm{tr}(\mathrm{sym}\nabla u) \right) I +k_{3}^{\prime
}\mathrm{sym}\nabla u 
\end{array}
\end{aligned}
\end{equation*}  
By setting
$ \xi = \lambda + \mu$, $\alpha =k_3 + \frac{1}{2} k'_3$,
$\zeta = k_2 + k'_2$, $\gamma = k_1 + k_2 - k'_2$, $\chi
=\frac{1}{2} k'_3\;,$
and choosing $\beta=b=0$, the local balances of momentum and microstructural actions reduce to
\begin{equation} \label{Momentum}
\rho \ddot{u}= \mu \Delta u+(\lambda+\mu)\nabla\mathrm{div}\,u+\chi \Delta \nu + \alpha \nabla\mathrm{div}\,\nu \;,
\end{equation}
and a balance of microstructural actions given by
\begin{equation} \label{Diffusion}
\varsigma\dot{\nu} + k_0 \nu = \chi \Delta u+\alpha \nabla\mathrm{div}\,u+\zeta \Delta \nu + \gamma \nabla\mathrm{div}\,\nu\;.
\end{equation}
The analyses reported in what follows refer to them.

\section{Waves with time-dependent amplitude}

Assume
\begin{equation}
u(x,t) =\mathrm{Re}\{ \mathbf u(t) e^{i\mathbf k\cdot \mathbf x}\} \quad \text{and} \quad  \nu(x,t) =\mathrm{Re}\{  \mathbf \nu(t) e^{i \mathbf k \cdot \mathbf x}\}\;,
\label{onda stazionaria campi di spostamento it}
\end{equation}
where, we repeat, $\mathbf k = k \mathbf{n}$ is a constant vector, the wave amplitudes $\mathbf u, \mathbf \nu \in \mathbb C^3$ depend on time, and $\mathbf{x}=x-x_{0}$, with $x_{0}$ an arbitrary point.

We consider various possible choices for the phason self-action in order to evaluate relevant effects.

\subsection{In the absence of a self-action}

If $\mathbf{z}=0$, by inserting expressions \eqref{onda stazionaria campi di spostamento it} into the balance equations, we get
\begin{equation}
\ddot{\mathbf u} = -k^2\bigg(\frac{\mu}{\rho} I + \frac{\lambda+\mu}{\rho}(\mathbf{n}\otimes \mathbf{n} )\bigg) \mathbf u -\frac{k^2}{\rho} \bigg(\chi I + \alpha (\mathbf{n}\otimes \mathbf{n} ) \bigg)\mathbf \nu
\label{OStazio_no Autoazione_ bilancio forze}
\end{equation}
and
\begin{equation}
\bigg(\zeta I + \gamma(\mathbf{n}\otimes \mathbf{n} )\bigg) \mathbf \nu = -\bigg(\chi I + \alpha (\mathbf{n}\otimes \mathbf{n} )\bigg)\mathbf u\;.
\label{OStazio_no Autoazione_ bilancio microstrutt}
\end{equation}
By substituting \eqref{OStazio_no Autoazione_ bilancio microstrutt} into \eqref{OStazio_no Autoazione_ bilancio forze} we get
\begin{equation}
\ddot{\mathbf u} = \mathbb K \mathbf u\;,
\label{OStazio_no Autoazione_ bilancio forze e microstrutt insieme}
\end{equation}
where
\begin{equation}\label{K-no-self}
\mathbb{K}:=-\frac{k^2}{\rho}\bigg[\left(\mu-\frac{\chi^2}{\zeta}\right)I +\left(\lambda+\mu-\frac{(\alpha+\chi)^2}{\zeta+\gamma}+\frac{\chi^2}{\zeta}\right)\mathbf{n}\otimes\mathbf{n}\bigg]\;.
\end{equation}

Thus, by defining
\begin{equation}\label{Absence of self-action}
\mathbf{q}:=\begin{bmatrix}
\mathbf u \\ \mathbf{v}
\end{bmatrix} \in \mathbb{R}^6\;,\qquad A:=\begin{bmatrix}
\mathbf{0} &&&& I \\ \mathbb{K} &&&& \mathbf{0}
\end{bmatrix}\in \mathbb{R}^{6\times 6}\;,
\end{equation} 
where $\mathbf{v}:=\dot{\mathbf u}$, we rewrite \eqref{OStazio_no Autoazione_ bilancio forze e microstrutt insieme} as
\begin{equation}
\dot{\mathbf q}=A\mathbf{q}\;.
\label{OndaStazio: corpo £££}
\end{equation}

We compute
\[AA^\top=\begin{bmatrix}
\mathbf{0} &&&& I \\ \mathbb{K} &&&& \mathbf{0}
\end{bmatrix}\begin{bmatrix}
\mathbf{0} &&&& \mathbb{K}^\top\\ I  &&&& \mathbf{0}
\end{bmatrix}=\begin{bmatrix}
I &&&& \mathbf{0} \\ \mathbf{0}&&&&\mathbb{K}\mathbb{K}^\top  
\end{bmatrix}\;,\]
while
\[A^\top A=\begin{bmatrix}
\mathbf{0} &&&& \mathbb{K}^\top\\ I  &&&& \mathbf{0}
\end{bmatrix}\begin{bmatrix}
\mathbf{0} &&&& I \\ \mathbb{K} &&&& \mathbf{0}
\end{bmatrix}=\begin{bmatrix}
\mathbb{K}^\top \mathbb{K}  &&&& \mathbf{0}\\ \mathbf{0} &&&& I   
\end{bmatrix}\;;\]
therefore, $A$ is \emph{non-normal} (normality being defined by the equality $AA^\top= A^\top A$ when $A\in \mathbb R^{n\times n}$ or by $AA^\ast= A^\ast A$, when $A\in \mathbb C^{n\times n}$).

Nonetheless, even if $A$ is non-normal, the stability of the system is completely described by its eigenvalues since
\begin{equation}
A^2=\begin{bmatrix}
\mathbf{0} &&&& I \\ \mathbb{K} &&&& \mathbf{0}
\end{bmatrix}\begin{bmatrix}
\mathbf{0} &&&& I \\ \mathbb{K} &&&& \mathbf{0}
\end{bmatrix}=\begin{bmatrix}
\mathbb{K} &&&& \mathbf{0} \\ \mathbf{0} &&&& \mathbb{K}
\end{bmatrix}\;.
\end{equation}
Therefore, the spectrum of $A$ is given by the square roots of the eigenvalues of the symmetric matrix $\mathbb{K}$, namely
\[\sigma(A)=\bigcup_{\lambda\in\sigma(\mathbb K)} \pm\sqrt{\lambda}\;.\]
Hence, the eigenvalues of $A$ are either real or purely imaginary. In particular the eigenvalues of $A$ are real, both positive and negative, when $\mathbb{K}$ is positive definite; imaginary when $\mathbb{K}$ is negative definite and real and imaginary when $\mathbb{K}$ is indefinite. 
Moreover, the spectrum of $A^2$ is stable under (symmetric) perturbations in $\mathbb{K}$ 
thanks to Weyl's theorem \cite{HJ13}, and thus the eigenvalues provide reliable information on the stability
of the system.

\subsection{In the presence of a conservative self-action}

Le us consider $\mathbf z = k_0 \nu$. The balance equations still reduce to a system of the form
\begin{equation}
\dot{\mathbf q}=A\mathbf{q}\;,
\label{OndaStazio: corpo comple z=k0 nu}
\end{equation}
with
\begin{equation} \label{A-conservative}
\mathbf{q}:=\begin{bmatrix}
\mathbf u \\ \mathbf{v}
\end{bmatrix}\quad \text{and} \quad A:=\begin{bmatrix}
\mathbf{0} &&&& I \\ \mathbb{K} &&&& \mathbf{0}
\end{bmatrix}\in \mathbb{R}^{6\times6}\;,
\end{equation}
but now
\begin{equation} \label{K-conservative}
\mathbb{K}:=-\frac{k^2}{\rho}\bigg[\left(\mu-\frac{\chi^2k^2}{k_0+\zeta k^2}\right)I +\bigg(\lambda+\mu-k^2\left(\frac{(\alpha+\chi)^2}{k_0+k^2(\zeta+\gamma)}-\frac{\chi^2}{k_0+\zeta k^2}\right)\bigg)\mathbf{n}\otimes\mathbf{n}\bigg]\;.
\end{equation}
As in the previous case, all the eigenvalues of $A$ are either real or purely imaginary. In order to analyze the system stability it suffices to compute the spectrum of the symmetric matrix $\mathbb{K}$: if $\mathbb{K}$ is either indefinite or negative definite, the system is unstable. Once again, the eigenvalues of $A$ are stable under small symmetric perturbations on $\mathbb{K}$. 

\subsection{A purely dissipative self-action}

Assume that $\mathbf{z}$ is endowed only with a dissipative component driving diffusion, namely it is of the form $\mathbf z = \varsigma \dot{\nu}$, with $\varsigma$ a positive constant, as chosen in Section \ref{Se4}. From the balance equations we obtain
\begin{equation}
\ddot{\mathbf u} = -k^2\bigg(\frac{\mu}{\rho} I + \frac{\lambda+\mu}{\rho}(\mathbf{n}\otimes \mathbf{n} \mathbf)\bigg) \mathbf u -\frac{k^2}{\rho} \bigg(\chi I + \alpha (\mathbf{n}\otimes \mathbf{n} \mathbf) \bigg)\mathbf \nu 
\label{eq: OndeStazio z completa bilancio macrostruttura}
\end{equation}
and
\begin{equation}
\dot{\mathbf \nu} = -\frac{k^2}{\varsigma} \bigg(\chi I + \alpha (\mathbf{n}\otimes \mathbf{n} \mathbf)\bigg)\mathbf u -k^2 \bigg(\frac{\zeta}{\varsigma} I + \frac{\gamma}{\varsigma} (\mathbf{n}\otimes \mathbf{n} \mathbf)\bigg) \mathbf \nu\;,
\label{eq: OndeStazio z completa bilancio microstruttura}
\end{equation}
which can be rewritten as
\begin{equation}
\ddot{\mathbf u} = \mathbb K_1 \mathbf u + \frac{1}{\rho}\mathbb K_2 \mathbf \nu \;, \qquad 
\dot{\mathbf \nu} = \frac{1}{\varsigma}\mathbb K_2 \mathbf u +\mathbb K_3 \mathbf \nu\;,
\label{onde stazionarie, sistema dopo aver definito K1,2,3}
\end{equation}
where
\begin{equation} \label{K1-dissipative}
\mathbb K_1 := -k^2\bigg(\frac{\mu}{\rho} I + \frac{\lambda+\mu}{\rho}(\mathbf{n}\otimes \mathbf{n} \mathbf)\bigg) \;,
\end{equation}
\begin{equation} \label{K2-dissipative}
\mathbb K_2 : = -k^2 \bigg(\chi I + \alpha (\mathbf{n}\otimes \mathbf{n} \mathbf) \bigg) \;,
\end{equation}
\begin{equation} \label{K3-dissipative}
\mathbb K_3:= -k^2 \bigg(\frac{\zeta}{\varsigma} I + \frac{\gamma}{\varsigma} (\mathbf{n}\otimes \mathbf{n} \mathbf)\bigg)\;.
\end{equation}

By setting
\begin{equation} \label{A-dissipative}
\mathbf q := \begin{bmatrix}
\mathbf u \\ \mathbf v \\ \mathbf \nu
\end{bmatrix}\in \mathbb R^9\;, \quad A : = \begin{bmatrix}
\mathbf{0} & I &  \mathbf{0} \\
\mathbb K_1 & \mathbf{0} & \frac{1}{\rho} \mathbb K_2 \\ 
\frac{1}{\varsigma} \mathbb K_2& \mathbf{0} & \mathbb K_3
\end{bmatrix}\in \mathbb{R}^{9\times 9}\;,
\end{equation}
where, as above, $\mathbf{v}:=\dot{\mathbf u}$, we rewrite system \eqref{onde stazionarie, sistema dopo aver definito K1,2,3} in the standard form
\begin{equation}
\dot{\mathbf q} = A \mathbf{q}\;.
\label{onde stazionarie q.=Aq}
\end{equation}
The matrix $A$ is non-normal because
\[AA^\top=\begin{bmatrix}
\mathbf{0} & I &  \mathbf{0} \\
\mathbb K_1 & \mathbf{0} & \frac{1}{\rho} \mathbb K_2 \\ 
\frac{1}{\varsigma} \mathbb K_2& \mathbf{0} & \mathbb K_3
\end{bmatrix}\begin{bmatrix}
\mathbf{0} & \mathbb K_1^\top & \frac{1}{\varsigma} \mathbb K_2^\top \\ I &  \mathbf{0} & \mathbf{0} \\ \mathbf{0} & \frac{1}{\rho} \mathbb K_2^\top & 
\mathbb K_3^\top
\end{bmatrix}=\begin{bmatrix}
I & 0 &0 \\ 0 & \mathbb K_1\mathbb K_1^\top+\frac{1}{\rho^2}\mathbb K_2\mathbb K_2^\top & \frac{1}{\varsigma} \mathbb K_1\mathbb K_2^\top + \frac{1}{\rho} \mathbb K_2\mathbb K_3^\top \\ 0 &  \frac{1}{\varsigma} \mathbb K_2\mathbb K_1^\top + \frac{1}{\rho} \mathbb K_3\mathbb K_2^\top & \frac{1}{\varsigma^2}\mathbb K_2\mathbb K_2^\top+ \mathbb K_3\mathbb K_3^\top
\end{bmatrix}\]
and
\[A^\top A=\begin{bmatrix}
\mathbf{0} & \mathbb K_1^\top & \frac{1}{\varsigma} \mathbb K_2^\top \\ I &  \mathbf{0} & \mathbf{0} \\ \mathbf{0} & \frac{1}{\rho} \mathbb K_2^\top & 
\mathbb K_3^\top
\end{bmatrix}\begin{bmatrix}
\mathbf{0} & I &  \mathbf{0} \\
\mathbb K_1 & \mathbf{0} & \frac{1}{\rho} \mathbb K_2 \\ 
\frac{1}{\varsigma} \mathbb K_2& \mathbf{0} & \mathbb K_3
\end{bmatrix}=\begin{bmatrix}
\mathbb K_1^\top\mathbb K_1+\frac{1}{\varsigma^2}\mathbb K_2^\top\mathbb K_2 & 0 &\frac{1}{\rho} \mathbb K_1^\top\mathbb K_2+\frac{1}{\varsigma}\mathbb K_2^\top\mathbb K_3 \\ 0 & I & 0 \\ \frac{1}{\rho} \mathbb K_2^\top\mathbb K_1+\frac{1}{\varsigma}\mathbb K_3^\top\mathbb K_2 & 0 & \frac{1}{\rho^2}\mathbb K_2^\top\mathbb K_2+\mathbb K_3^\top\mathbb K_3
\end{bmatrix}\;.\]
At variance with the previous two cases, the non-normality in $A$ is now genuine, and we can no longer be sure that small perturbations on $\mathbb K_i$ will result in small changes in the spectrum of $A$. In other words, the spectrum of $A$ could be highly sensitive to small changes in the parameters, making stability predictions based on the eigenvalues alone unreliable, especially in the presence of uncertainties or rounding errors in computed quantities. To assess  stability we need to have a look at the pseudospectrum of $A$.

\subsection{Simultaneous presence of conservative and dissipative components in the self-action}

Finally, we consider $\mathbf z = k_0  \nu + \varsigma \dot{\nu}$, so that, with previous notations, the matrix $A$ in the system
\begin{equation}
\dot{\mathbf q} = A \mathbf{q}\;,
\label{onde stazionarie q.=Aq z completa}
\end{equation}
becomes
\begin{equation} \label{A-complete}
A : = \begin{bmatrix}
\mathbf{0} & I &  \mathbf{0} \\
\mathbb K_1 & \mathbf{0} & \frac{1}{\rho} \mathbb K_2 \\ 
\frac{1}{\varsigma} \mathbb K_2& \mathbf{0} & \mathbb K_3
\end{bmatrix}\;,
\end{equation}
with
\begin{equation} \label{K1-complete}
\mathbb K_1 := -k^2\bigg(\frac{\mu}{\rho} I + \frac{\lambda+\mu}{\rho}(\mathbf{n}\otimes \mathbf{n} \mathbf)\bigg) \;,
\end{equation}
\begin{equation} \label{K2-complete}
\mathbb K_2 : = -k^2 \bigg(\chi I + \alpha (\mathbf{n}\otimes \mathbf{n} \mathbf) \bigg) \;,
\end{equation}
\begin{equation} \label{K3-complete}
\mathbb K_3:= -k^2 \bigg(\frac{\zeta + \frac{k_0}{k^2}}{\varsigma} I + \frac{\gamma}{\varsigma} (\mathbf{n}\otimes \mathbf{n} \mathbf)\bigg)\;.
\end{equation}

Similar considerations to those made in the previous case apply here.

\section{Pseudospectrum: general properties}

\subsection{Reasons for calling upon the pseudospectrum}

For a system of the form \eqref{onde stazionarie q.=Aq z completa}, information on the asymptotic behavior of solutions are given by the spectrum of $A$. Specifically, as it is well-known, when all the eigenvalues of $A$ have a negative or null real part the system is said to be stable. In contrast, an amplified wave to infinity is associated with an eigenvalue of positive real part; in this case the system is said to be unstable. A parametric analysis may reveal possible transitions from stability to instability and vice versa.

The analysis along these lines is straightforward when $A$ is a normal matrix. The eigenvalues indicate characteristic times of exponentially relaxing modes whose linear combination is a solution to the considered system.
When small perturbations are accounted for, small is the variation of the $A$ spectrum.
At variance, when $A$ is non-normal, as in the cases we tackle here, small perturbations can have significant effects that may induce instability. More specifically, for matrices far from being normal the standard spectral analysis shows possible failures in the evaluation of resonance,
asymptotic behavior, diagonalization, geometric characterization \cite{TE}. Moreover, little or no information on the transient behavior and rate of convergence to steady-state can be gleaned in the highly non-normal case from the eigenvalues alone \cite{B21}.

A way to manage such aspects is to consider an extended notion of spectrum, what is called a $\varepsilon$-\emph{pseudospectrum} \cite{TE}, \cite{Ghe14}, \cite{MGWN06}, \cite{Sj19}.

If there exists a small $\varepsilon$ such that the $\varepsilon$-pseudospectrum trespasses in the half plane with positive real part, the system has to be treated as unstable.

\subsection{Scalar measures of departure from normality}
Let $A \in \mathbb C^{n\times n}$ be a non-normal matrix and let $\mathfrak{K}$ be the set of $n\times n$ normal matrices. 
A measure of departure from normality is a real positive valued function on the space of matrices $\dep: \mathbb C^{n\times n}\rightarrow \mathbb R^+$ such that 
$\dep(A)=0$ if and only if $A\in \mathfrak{K}$.\\ 
The most natural measure of non-normality is given by the distance between $A$ and the set $\mathfrak{K}$, namely 
\begin{equation}
    {\dist}(A,\mathfrak{K})=\inf_{N\in \mathfrak{K}}\| A-N\| \;,
\end{equation}
in some matrix norm $\|\cdot\|$. In the following, we restrict ourselves to the Frobenius norm $\| A \|_F = (\sum_{i,j=1}^n |a_{ij}|^2)^{1/2}$.
In the literature, many other scalar measures of departure from normality have been introduced, which are easier and cheaper to compute (see, e.g., \cite{Els87} for a list).\\
Assume that $A$ is diagonalizable and not necessarily normal. Indicate by $V$ a matrix of eigenvectors $\mathbf{v}$ of $A$, namely
\[ V=\begin{bmatrix}
\mathbf v_1 & \dots & \mathbf v_n
\end{bmatrix}\in\mathbb{C}^{n\times n}\;,\] 
where each column $\mathbf v_j$ is normalized by $\| \mathbf v_j\|_2=1$. The number
\begin{equation}
\kappa_2(V):=\| V\|_2 \| V^{-1}\|_2 = \frac{s_{\max}(V)}{s_{\min}(V)}\;,
\end{equation}
where $s_{\max}(V)$ and $s_{\min}(V)$ are the largest and smallest singular values of $V$, is called the \emph{spectral condition number} of $V$.
In general, $1\leq \kappa_2(V)<\infty$; specifically, if $A$ is a normal matrix, we have $\kappa_2(V)=1$.
It is known that if $A$ is far from normal, then $\kappa_2 (V)$ will be large.  While the converse is not true in general, an ill-conditioned $V$ often signals that at least one of the eigenvalues of $A$ will be sensitive to small perturbations. Hence, one can use the spectral condition number of $V$, $\kappa_2(V)\gg 1$, as a measure of non-normality, albeit with caution. \\
If $A$ is normal, the commutator $[A,A^*]=AA^*-A^*A$ vanishes. Thus, one more natural measure of non-normality is given by the norm of the commutator between the matrix $A$ and its conjugate transpose $A^{\ast}$, divided by the square of the norm of $A$, to make the measure scale invariant, namely
\begin{equation}\label{Commutator Departure from normality}
\dep^c(A):=\frac{\| A^{\ast}A-AA^{\ast}\|_2}{\| A \|_2^2}\;.
\end{equation}
It has been proved (see \cite{Eberlein65}) that $0\leq \dep^c(A) \leq \sqrt{2}$. Of course, $\dep^c(A)=0$ if and only if $A$ is normal. Another measure of departure from normality, introduced by Henrici in \cite{Hen62}, is based on the Schur decomposition. Let $A\in \mathbb C^{n\times n}$ be any matrix and let $A=U(D+T)U^*$ be the Schur decomposition of $A$, where $D=\diag(\lambda_1,\dots,\lambda_n)$ is a diagonal matrix whose main diagonal is made up of the eigenvalues of $A$ and $T$ is a strictly upper triangular matrix. A classical result is that $T=0$ if and only if $A$ is normal. Therefore, Henrici's departure from normality is defined as 
\begin{equation}\label{Henrici's departure from normality}
    \dep^H(A):=\inf_{A=U(D+T)U^*}\| T \| \;,
\end{equation}
where the infimum is taken over all possible Schur decompositions. However, working in the Frobenius norm this quantity becomes independent on the particular decomposition: the quantity
\begin{equation}\label{Henrici's departure from normality Frobenius}
\dep^H_F(A)=\| T \|_F = \sqrt{\| A \|_F^2 - \| D\|_F^2}
\end{equation}
is the same for any decomposition $A=U(D+T)U^*$ of $A$. \\
Using Henrici's departure from normality it is possible to bound the distance of $A$ from the set of normal matrices $\mathfrak{K}$ both from above and below. Indeed, it holds that (see \cite{Els87} and \cite{Lasz94} for the proof)
\begin{equation} \label{Lower and upper bound departure}
    \frac{1}{\sqrt{n}} \dep^H_F(A)\leq {\dist}(A,\mathfrak{K}) \leq \dep^H_F(A)\;,
\end{equation}
where the distance is computed in the Frobenius norm. However, departure from normality does not imply necessarily eigenvalue instability under perturbations. The computation of the complex and of the real-structured $\varepsilon$-pseudospectra, recalled in the following section, for suitable values of $\varepsilon$ is a tool for detecting possible eigenvalue instabilities due to matrix non-normality.

\subsection{Definitions and a few properties}

Let $A$ be a $n\times n$ matrix with complex entries; in short, $A\in\mathbb C^{n\times n}$. The set $\sigma(A)$ of $A$ eigenvalues contains all $\lambda\in\mathbb{C}$ satisfying $\det(A-\lambda I)=0$, with $I$ the $n\times n$ unit matrix. The singular character of a matrix is however perturbation sensitive: if $A\in \mathbb{C}^{n\times n}$ is singular, there exists always $E\in \mathbb{C}^{n\times n}$, with $\| E \|_2\ll 1$ such that the matrix $A+E$ is non-singular. So, instead of asking whether $\lambda I-A$ is singular, we may look at the norm of its inverse, checking whether the value is sufficiently large. Of course, choosing the norm of its inverse matters; here we refer to the matrix norm induced by the Euclidean one.

We have three equivalent definitions of \emph{$\varepsilon$-pseudospectrum} \cite{TE}.

\begin{definition} [First version]
Given $A\in \mathbb C^{n\times n}$, take an arbitrary $\varepsilon >0$. The $\varepsilon$-pseudospectrum of $A$, indicated by $\sigma_{\varepsilon}(A)$, is the set of $\hat{\zeta}\in\mathbb C$ such that
\begin{equation}
\| (\hat{\zeta} I-A)^{-1}\|_2 > \frac{1}{\varepsilon}\;.
\label{prima definizione pseudospettro}
\end{equation}
\end{definition}

Here we use the convention 
$\|(\hat{\zeta} I - A)^{-1}\|_2 = \infty$ 
if $\hat{\zeta} I - A$ is singular.

\begin{definition} [Second version]
The $\varepsilon$-pseudospectrum $\sigma_{\varepsilon}(A)$ is the set of complex numbers $\zeta\in \mathbb C$ such that
\begin{equation}
\hat{\zeta}\in \sigma(A+E),
\label{seconda definizione pseudospettro}
\end{equation}
for some $E\in \mathbb{C}^{n\times n}$ with $\| E \|_2 < \varepsilon$.
\end{definition}

\begin{definition} [Third version]
The $\varepsilon$-pseudospectrum $\sigma_{\varepsilon}(A)$ is the set of complex number $\zeta\in\mathbb C$ such that
\begin{equation}
\|(\hat{\zeta} I-A)\mathbf{v}\|_2 < \epsilon
\label{terza definizione pseudospettro}
\end{equation}
for some $\mathbf{v}\in \mathbb{C}^{n\times n}$, with $\| \mathbf{v} \|_2=1$.
\end{definition}

The idea---we repeat---is thus to compute the spectra of matrices differing from $A$ by a perturbation given by matrices in a ball of radius $\varepsilon$, evaluating in this way \emph{how far the spectrum of $A$ can go under perturbations}.\\
Given $\varepsilon>0$, the pseudospectrum of $A\in \mathbb{C}^{n\times n}$, namely $\sigma_{\varepsilon}(A)$, referred to the Euclidan norm, satisfies what follows (see \cite{TE} for the pertinent proofs):
\begin{enumerate}
\item $\sigma_{\varepsilon}(A)$ is an open bounded non-empty set composed of at most $n$ connected subsets, each containing one or more eigenvalues of $A$.
\vskip6pt
\item $\sigma_{\varepsilon}(A^*)=\overline{\sigma_{\varepsilon}(A)}$, the conjugate of the pseudospectrum of $A$.
\vskip6pt
\item $\sigma_{\varepsilon}(A_1 \oplus A_2)=\sigma_{\varepsilon}(A_1)\cup\sigma_{\varepsilon}(A_2)$, where
\[A_1 \oplus A_2 = \begin{bmatrix}
A_1 & 0 \\ 0 & A_2
\end{bmatrix}\;.\]
\vskip6pt
\item $\forall \mathsf{h}\in \mathbb C$, $\sigma_{\varepsilon}(A+\mathsf{h}I )= \mathsf{h} +\sigma_{\varepsilon}(A)$.
\vskip6pt
\item $\forall \mathsf{h}\in \mathbb C$, $\mathsf{h}\not =0$, we get
\[ \sigma_{|\mathsf{h}|\varepsilon}(\mathsf{h}A)= \mathsf{h}\sigma_{\varepsilon}(A)\;.\]
In particular, for $\mathsf{h}=2$ we have
\[\sigma_{\varepsilon}(2A)= 2\sigma_{\frac{\varepsilon}{2}}(A)\;.\]
\end{enumerate}

\subsection{Structured pseudospectra}

The matrices \eqref{A-dissipative} and \eqref{A-complete} have a block-type structure, namely
\begin{equation}\label{A-Block matrix}
A= \begin{bmatrix}
\mathbf{0} &&& I &&&  \mathbf{0} \\
A_1 &&& \mathbf{0} &&& \frac{1}{\rho}A_2 \\ 
\frac{1}{\varsigma}A_2 &&& \mathbf{0} &&& A_3
\end{bmatrix}\in \mathbb{R}^{9\times 9}\;.
\end{equation}
 where the $A_i$ blocks are symmetric $3\times 3$ matrices. 
In view of this structure, it may be appropriate to perturb only the blocks that are different from the identity and the null matrix.
The notion of \emph{structured pseudospectrum} enters thus into play \cite{Rie94}, \cite{TH01}, \cite{TE}.

\begin{definition}
The structured $\varepsilon$-pseudospectrum of a matrix $A\in \mathbb R^{N\times N}$ of the form \eqref{A-Block matrix} is defined as
\begin{equation}
\sigma_\varepsilon^{Str}(A)= \bigcup_{\|E\|_2 < \varepsilon} \sigma(A+E) \;,
\end{equation}
 with
\begin{equation} \label{Perturbations E well scaled}
    E : = \begin{bmatrix}
\mathbf{0} & \mathbf{0} &  \mathbf{0} \\
\frac{1}{\rho}E_1 & \mathbf{0} & \frac{1}{\rho}E_2 \\ 
\frac{1}{\varsigma} E_2& \mathbf{0} & \frac{1}{\varsigma}E_3
\end{bmatrix}\in \mathbb R^{9\times 9}\;,
\end{equation}
 with $E_i \in Sym(\mathbb R^3,\mathbb R^3)$ and $\| E \|_2<\varepsilon$. 
\end{definition}
The structured $\varepsilon$-pseudospectrum is of course a subset of the complex $\varepsilon$-pseudospectrum, $\sigma_\varepsilon^{Str}(A) \subset \sigma_\varepsilon(A)$. \\
In our particular case the second and the third blocks in the matrices defined in equations \eqref{A-dissipative} and \eqref{A-complete} have different scales. To consider the circumstance that events described by $\nu$ occur at a scale different from the macroscopic one, we consider perturbations of the form \eqref{Perturbations E well scaled}; this way we are consistent with the physics discussed.

The complex pseudospectrum is computed considering the close relation between the eigenvalues of the perturbed matrices and the resolvent norm. In the case of real structured perturbations, such a relation does not exist. Thus, we proceed by computing the spectrum of $400$ perturbed matrices $A+E$, where $E$ has the structure given by \eqref{Perturbations E well scaled}.
\\ \\

\subsection{On the values of the admissible perturbations}
We look at perturbations of material parameters due to possible uncertainty in measurements of such parameters. Assume, for example, that the value of the coupling parameter $\chi$ is known up to an error of the order of $5\%$. Thus, we can write $\chi=\chi_0 \pm \tilde{\chi}$, with $\tilde{\chi}=0.05\chi_0$ and $\chi_0$ the nominal value adopted.   
Since the matrix $A$ has linear dependence with respect to some of the material parameters (e.g. $\chi$ and the Lamè constants), the idea is to split $A$ into the sum of two matrices $A=A_0+\tilde{A}$, where $A_0$ is evaluated with the nominal values of the material parameters while $\tilde{A}$ is computed over the perturbations. Then we consider the 2-norm of the matrix $\tilde{A}$ as the maximum value of $\varepsilon$ physically meaningful for the computation of the structured $\varepsilon$-pseudospectrum. In the following, we say that a certain perturbation matrix $E$ with 2-norm equal to $\varepsilon$ corresponds to a perturbation of the order of $5\%$ on the values of the material parameters when the matrix $\tilde{A}$, evaluated with the perturbed values of the constitutive coefficients, has 2-norm equal to $\varepsilon$.

\section{Pseudospectrum in the quasicrystal dynamics: implications}

We refer explicit computations to quasicrystals, as already anticipated. The software used  for the complex pseudospectrum is the one developed by Wright \cite{Wri}.
Data considered are as follows:
\[\lambda=85\,\mathrm{GPa}\;,\quad \mu=65\,\mathrm{GPa}\;,\quad\zeta=0.044\,\mathrm{GPa}\;,\quad\gamma=0.0198\,\mathrm{GPa}\;.\]
Mass density is $\rho=5100\,\mathrm{kg\,m^{-3}}$ \cite{Wa03}, \cite{Let}. Values of the coupling parameters $\chi$ and $\alpha$ as well as those of $k_0$ and $\varsigma$, which enter the self-action, are unknown. However, we know a theoretical range for the coupling parameter $\chi$, which is $[0,1.5]\,\mathrm{GPa}$, as established in \cite{RM11}. We also take the friction coefficient $\varsigma$ of the form $\varsigma= e^{\phi} \,\mathrm{Pa\cdot s/m^2}$, with $\phi\in[15,19]$.

The focus is on solutions of the form
\[ u = \mathrm{Re}\{\mathbf u(t) e^{i(\mathbf k \cdot \mathbf x)}\}\;,\quad  \nu = \mathrm{Re}\{\mathbf \nu(t) e^{i(\mathbf k \cdot \mathbf x)}\}\;,\]
where the time-dependent amplitudes $\mathbf u$ and $\mathbf \nu$ are complex. \\

The entries in $A$ are in practice always known only approximately. Among the sources of uncertainty in the matrix entries there are rounding errors due to finite precision computer arithmetic and those due to the experimental evaluation of physical parameters.

In the following analysis we consider both complex and structured pseudospectra with different purposes: 
\begin{itemize}
    \item The complex pseudospectra $\sigma_\varepsilon(A)$ are a useful tool, e.g., for understanding if the eigenvalues are stable under perturbations due to operations in finite precision arithmetic. 
    For this type of stability analysis, the values of $\varepsilon$ considered are thus of the order of machine precision. 
    \item The structured pseudospectra $\sigma^{Str}_\varepsilon(A)$ are used to understand if an uncertainty on the nominal values of the materials parameters may cause a transition towards instability before the one predicted by the eigenvalues. Thus, the order of the perturbations must have a physical meaning. In particular $\varepsilon$ must correspond to a certain variation (e.g., due to experimental uncertainty) of the values of the constitutive coefficients. Since, as shown in the results, the system becomes unstable for large values of the coupling parameter $\chi$, in the following we choose to adopt values of $\varepsilon$ corresponding to a variation of the order of $5\%$ of $\chi$. If any of the eigenvalues of perturbed matrices $A+E$, where $E$ is of the form \eqref{Perturbations E well scaled}, have positive real part, we conclude that the system must be regarded as unstable. 
\end{itemize}

\subsection{No phason self-action}

Here we consider the matrix $A$ in \eqref{Absence of self-action}
with $\mathbb{K}$ given by \eqref{K-no-self}.
Table \ref{tab: Risult.Onde staz: no autoazione; varia chi} lists the eigenvalues of $A$ as $\chi$ varies and $\alpha =0$ GPa.

For $\chi<1.6912$ all the eigenvalues of $A$ are purely imaginary:  the system undergoes infinitely many oscillations without damping.

When $\chi$ grows, the eigenvalues of $A$ that correspond to transversal waves migrate towards the real axis. Also, for $\chi \geq 1.6912$ GPa two coincident eigenvalues become real and positive, meaning that for such values of the parameters $\mathbb{K}$ is indefinite. Indeed, for $\chi=1.6912$ GPa and $\alpha=0$ GPa we have
\[\sigma(\mathbb{K})=\{-3.34e+07\;,\;701.6043\;,\;701.6043\}\;.\]
Thus, for $\chi \geq 1.6912$ GPa \emph{instability occurs}.
The limit is in agreement with what has been theoretically foreseen in \cite{RM11}. 
\begin{table}
\caption{No phason self-action: spectrum of $A$ for some values of $\chi$.}
\label{tab: Risult.Onde staz: no autoazione; varia chi}
\begin{center}
	\begin{tabular}{ |c|c|c|c|c|c| } 
		\hline
		\multicolumn{6}{|c|}{$\alpha=0$ GPa} \\
		\hline
		$\chi$ $[\mathrm{GPa}]$ & $0.05$ & $0.5$ & $1.6$ & $1.6912$ & $1.7$ \\
		\hline
		\multirow{6}{4em}{$\sigma(A)$ $[\mathrm{rad/s}]$ } 
		& $6492.24\,i$ & $6433.39\,i$ & $5855.69\,i$ & $5776.39\,i$ & $5768.44\,i$ \\
		& $6492.24\,i$ & $-6433.39\,i$ & $-5855.69\,i$ & $-5776.39\,i$ & $-5768.44\,i$ \\
		& $3568.47\,i$ & $3410.43\,i$ & $1156.24\,i$ & $\textcolor{red}{+}26.49 $ & $\textcolor{red}{+}365.64 $ \\
		& $-3568.47\,i$ & $-3410.43\,i$ & $-1156.24\,i$ & $-26.49 $ & $-365.64 $ \\
		& $3568.47\,i$ & $3410.43\,i$ & $1156.24\,i$ & $\textcolor{red}{+}26.49 $ & $\textcolor{red}{+}365.64 $ \\
		& $-3568.47\,i$ & $-3410.43\,i$ & $-1156.24\,i$ & $-26.49 $ & $-365.64 $ \\
		\hline
	\end{tabular}
\end{center}
\end{table}


\subsection{Phason self-action with purely conservative character}

When $\mathbf z=k_0\nu$ we refer to the matrix $A$ given by \eqref{A-conservative} for which $\mathbb{K}$ is given by formula \eqref{K-conservative}. 

Table \ref{tab: Risult.Onde staz: autoazione conservativa; varia k0} collects the eigenvalues of $A$ as $k_0$ varies, with $\chi=0.1$ GPa and $\alpha =0$ GPa. Given $\chi\in[0,\;1.6]$ GPa, stability is assured for any choice of $k_0$. For small $k_0$ the eigenvalues of $A$ are near those in the case $\mathbf{z}=0$; also, as $k_0$ grows, the eigenvalues tend to those of a simple body, intended in the traditional sense \cite{N58}, \cite{Tru77}, namely a body in which the microstructure representation in terms of phase fields is neglected and the stress depends only on the deformation gradient. The self-action is in essence a reaction to changes inside each material element, now considered as a system, rather than a black box collapsed at a point, as in the traditional description of continua. Thus, as $k_0\rightarrow \infty$ the inner system of each material element is more and more frozen so that in a sense it tends more and more to behave like a black box.

\begin{table}
\caption{Conservative self-action: spectrum of $A$ for some reasonable values of $k_0$.}
\label{tab: Risult.Onde staz: autoazione conservativa; varia k0}
\begin{center}
	\begin{tabular}{ |c|c|c|c|c| } 
		\hline
		\multicolumn{5}{|c|}{$\chi=0.1$ GPa, $\alpha=0$ GPa} \\
		\hline
		$k_0$ $[\mathrm{GPa/m^2}]$ & $0.001$ & $0.01$ & $0.1$  & $1$ \\
		\hline
		\multirow{6}{4em}{$\sigma(A)$ $[\mathrm{rad/s}]$ } 
		& $6492.81\,i$ & $ 6492.81\,i$ & $ 6492.82\,i$ & $ 6492.83\,i$\\
		& $-6492.81\,i$ & $ -6492.81\,i$ & $ -6492.82\,i$ & $ -6492.83\,i$\\
		& $3569.97\,i$ & $ 3569.98\,i$ & $ 3570.01\,i$ & $  3570.03\,i$\\
		& $-3569.97\,i$ & $ -3569.98\,i$ & $ -3570.01\,i$ & $ -3570.03\,i$\\
		& $3569.97\,i$ & $ 3569.98\,i$ & $ 3570.01\,i$ & $ 3570.03\,i$\\
		& $-3569.97\,i$ & $ -3569.98\,i$ & $ -3570.01\,i$ & $ -3570.03\,i$\\
		
		\hline
	\end{tabular}
\end{center}
\end{table}

However, as $k_0$ increases, the stability range for the coupling parameter $\chi$ increases, meaning that the system becomes unstable for higher values of $\chi$. Specifically, for $k_0=0.01$ $\mathrm{GPa/m^2}$, instability occurs when $\chi\geq 1.88$ GPa; $k_0=0.1$ $\mathrm{GPa/m^2}$ corresponds to an unstable state for $\chi\geq 3.06$ GPa; the value $k_0=1$ $\mathrm{GPa/m^2}$ is a threshold towards instability for $\chi\geq 8.24$ GPa.
In other words, the microstructural (phason in this case) self-action \emph{stabilizes the material}.


\subsection{Microstructural self-action with purely dissipative character}

When $\mathbf z=\varsigma \dot{\nu}$, we consider the matrix in formula \eqref{A-dissipative} with $\mathbb{K}_{i}$, $i=1,2,3$ given by formulas \eqref{K1-dissipative}, \eqref{K2-dissipative}, and \eqref{K3-dissipative}.\\ \\
 
\begin{figure}
\centering
\includegraphics[width=1\linewidth]{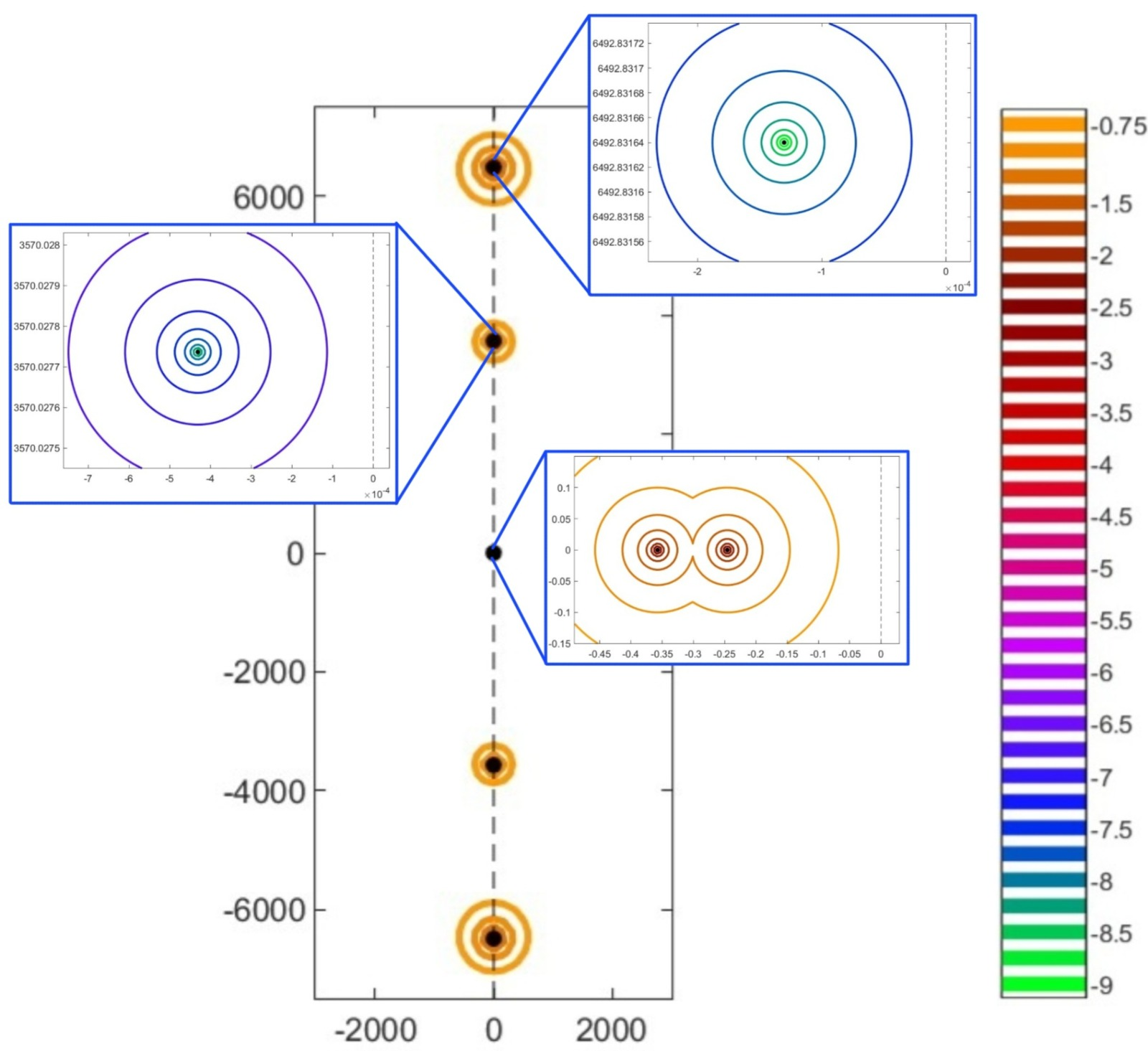}
\caption{ Complex $\varepsilon$-pseudospectrum of $A$ for $\chi=0.1$ GPa and $\phi=19$. The dashed line is the imaginary axis. }
\label{fig: Complex PS with zoom. Chi=0.5 GPa}
\end{figure}

For this matrix, the measures of departure from normality are $\kappa_2(V) = 6.50\cdot 10^3$, $\dep^c(A)=1$ (up to six digits) and $\dep_F^H=4.58\cdot 10^7$. Thus, from the bound \eqref{Lower and upper bound departure} we find
\begin{equation}
    1.52\cdot 10^7\leq {\dist}(A,\mathfrak{K}) \leq 4.48\cdot 10^7\;.
\end{equation} 
These measures confirms that the matrix $A$ is genuinely non-normal. Also, they remain approximately constant for different values of $\chi$ and $\varsigma$ in the ranges considered in this work. Figure \ref{fig: Complex PS with zoom. Chi=0.5 GPa} shows the boundary of the 2-norm complex pseudospectra of $A$ for $\chi=0.1$ GPa and $\phi=19$. The ticks on the color-bar on the right correspond to the 10-based logarithm of $\varepsilon$, i.e. the contour correspond to $\varepsilon=10^{-0.75},10^{-1.5}$, etc. For $\varepsilon \leq 10^{-6}$, $\sigma_\varepsilon(A)$ is completely contained in the left half part of the complex plane. Even though $A$ is genuinely non-normal, its eigenvalues are stable under perturbations due to operations in finite precision. This is true for any value of the material parameters $\chi$ and $\varsigma$ considered.

\begin{figure} [ht!]
\centering
\subfloat
{\includegraphics[width=0.49\textwidth]{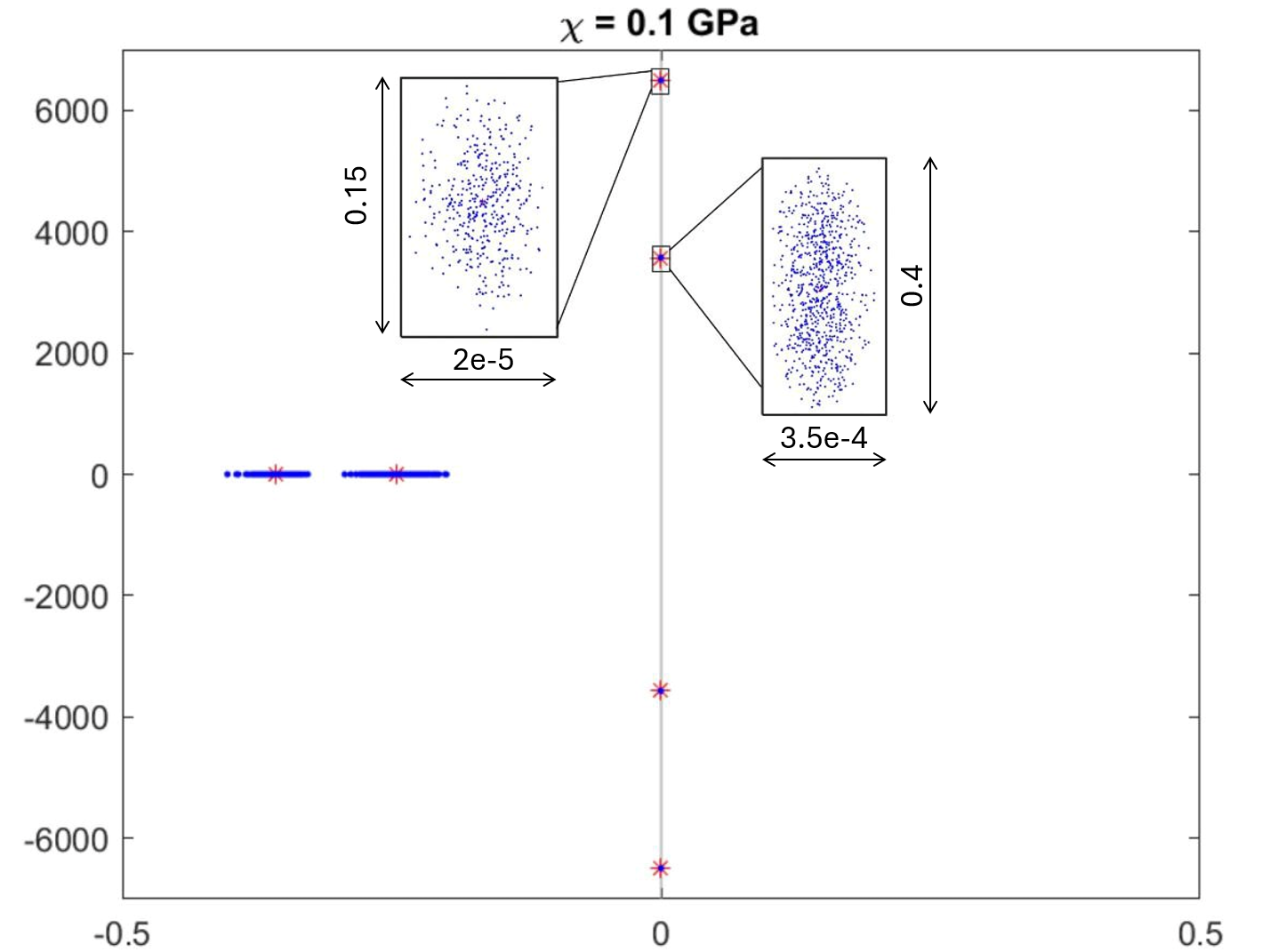}} \;
{\includegraphics[width=0.49\textwidth]{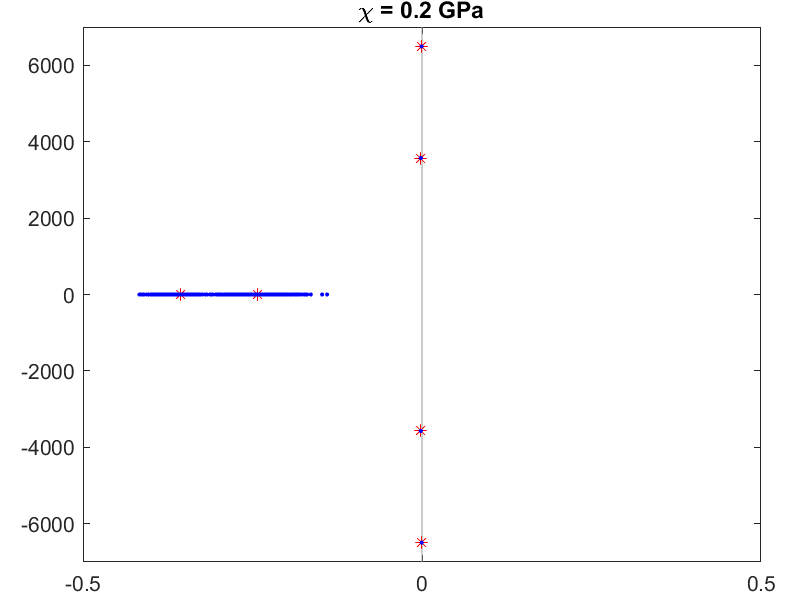}} \\
{\includegraphics[width=0.48\textwidth]{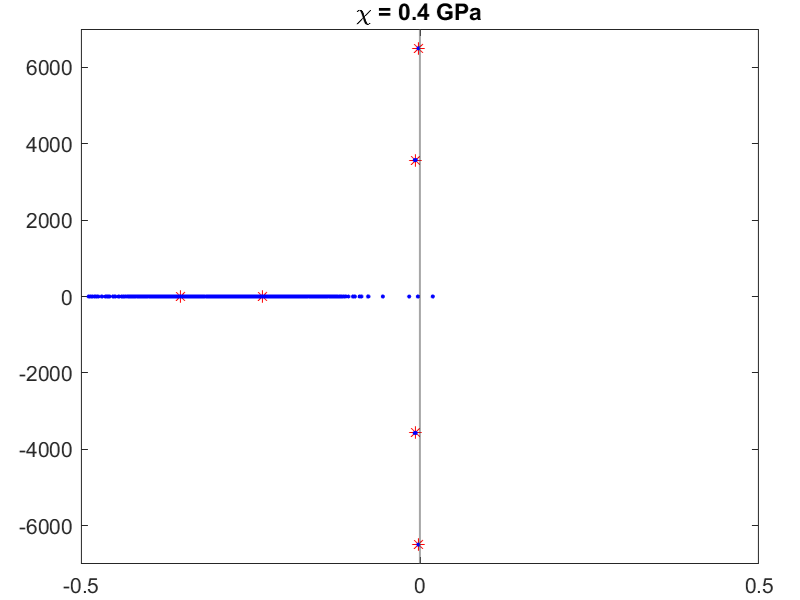}} \;
{\includegraphics[width=0.48\textwidth]{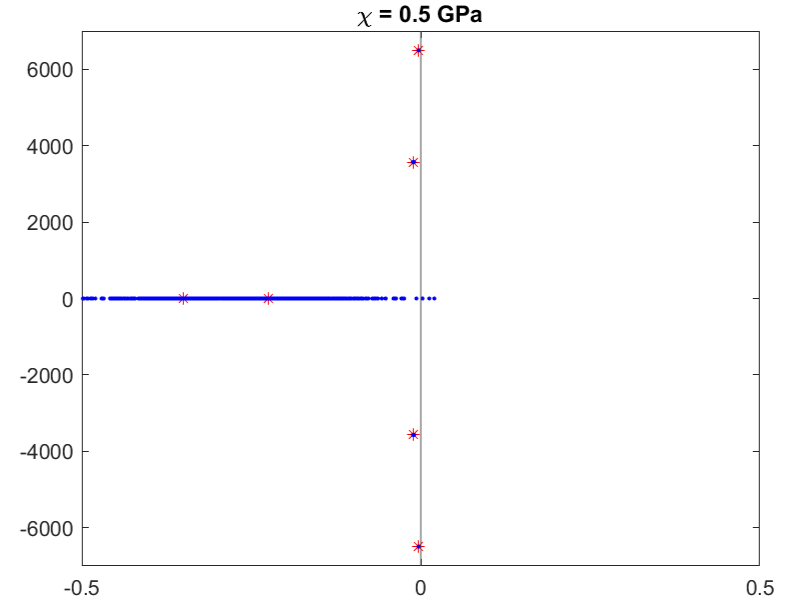}} 
\caption{Dissipative self-action: structured pseudospectrum of $A$ as $\chi$ varies. Superposition of eigenvalues of $400$ matrices $A+E$, where $E$ is a matrix with random entries and the structure given in \eqref{Perturbations E well scaled}.}
\label{fig: Autoazione dissipativa Pseudospettro di A al variare di chi}
\end{figure}

Table \ref{tab: Risult.Onde staz: autoazione dissipativa;} lists the eigenvalues of $A$ for $\alpha=0$ GPa, $\varsigma=0.178$ and $\mathrm{GPa\cdot s/m^2}$, as $\chi$ varies. When $\chi=1.7$ GPa the last two real eigenvalues are positive: \emph{instability occurs}. We do not find the same result by looking at the structured pseudospectrum in Figure \ref{fig: Autoazione dissipativa Pseudospettro di A al variare di chi} as $\chi$ varies, for $\varsigma=0.178\,\mathrm{GPa\cdot s/m^2}$, that is, $\phi=19$ and $\alpha=0$ GPa. For $\chi \geq 0.4$ GPa, the structured $\varepsilon$-pseudospectra, for $\varepsilon$ corresponding to $5\%$ of $\chi$, crosses the imaginary axis: the real eigenvalues of some perturbed matrices $A+E$ are positive. The complex eigenvalues of $A$ are instead stable under such a type of perturbation, as shown in the clouds in the first plot of Figure \ref{fig: Autoazione dissipativa Pseudospettro di A al variare di chi}. This behavior occurs for any value of the friction factor $\varsigma$ ($\phi$ varying from $15$ to $20$): when $\chi \geq 0.4$ GPa, the pseudospectra predict a transition towards instability of the system. We conclude that a dissipative component of the phason self-action does not change the (numerical) admissible range for the coupling parameter and, in this sense, it does not further stabilize the matter. Moreover, due to matrix non-normality, the analysis of the stability based on the spectrum $\sigma(A)$ does not  correctly predict the transition towards instability.

\begin{table}
\caption{Dissipative self-action: spectrum of $A$ for some values of $\chi$.}
\label{tab: Risult.Onde staz: autoazione dissipativa;}
\begin{center}
	\begin{tabular}{ |c|c|c|c|c| } 
		\hline
		\multicolumn{5}{|c|}{$\alpha=0$ GPa, $\phi=19 $} \\
		\hline
		$\chi$ $[\mathrm{GPa}]$ & $0.05$ & $0.5$ & $1.6$  & $1.7$ \\
		\hline
		\multirow{5}{4em}{$\sigma(A)$ $[\mathrm{rad/s}]$ } 
		& $-3\cdot 10^{-5}+6493\,i$ & $-3\cdot 10^{-3}+6493\,i$ & $-0.03+6493\,i$ & $-0.04+6493\,i$\\
		& $-3\cdot 10^{-5}-6493\,i$ & $-3\cdot 10^{-3}-6493\,i$ & $-0.03-6493\,i$ & $-0.04-6493\,i$\\
		& $-0.36$ & $-0.35$ & $-0.29$ & $-0.28$\\
		& $-1\cdot 10^{-4}+3570\,i$ & $-0.01+3570\,i$ & $-0.11+3570\,i$ & $-0.12+3570\,i$\\
		& $-1\cdot 10^{-4}-3570\,i$ & $-0.01-3570\,i$ & $-0.11-3570\,i$ & $-0.12-3570\,i$\\
		& $-1\cdot 10^{-4}+3570\,i$ &  $-0.01-3570\,i$  & $-0.11+3570\,i$ & $-0.12+3570\,i$\\
		& $-1\cdot 10^{-4}-3570\,i$ & $-0.01+3570\,i$ & $-0.11-3570\,i$ & $-0.12-3570\,i$\\
		& $-0.25$ & $-0.22$ & $-0.03$ & \textcolor{red}{+}$0.003$\\
		& $-0.25$ & $-0.22$ & $-0.03$ & \textcolor{red}{+}$0.003$\\
		\hline
	\end{tabular}
\end{center}
\end{table}

\begin{figure}
\centering
\subfloat
{\includegraphics[width=0.49\textwidth]{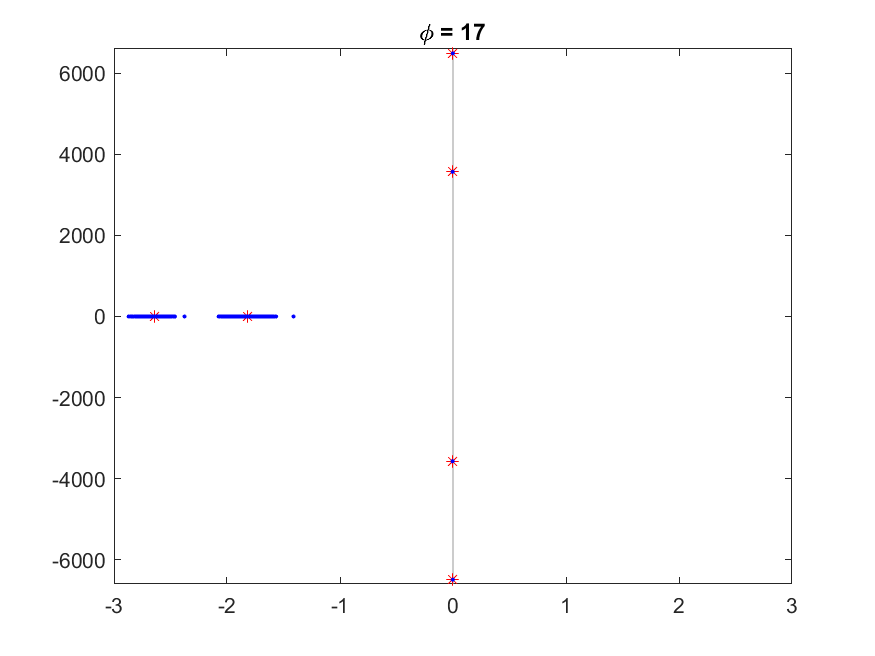}} \;
{\includegraphics[width=0.49\textwidth]{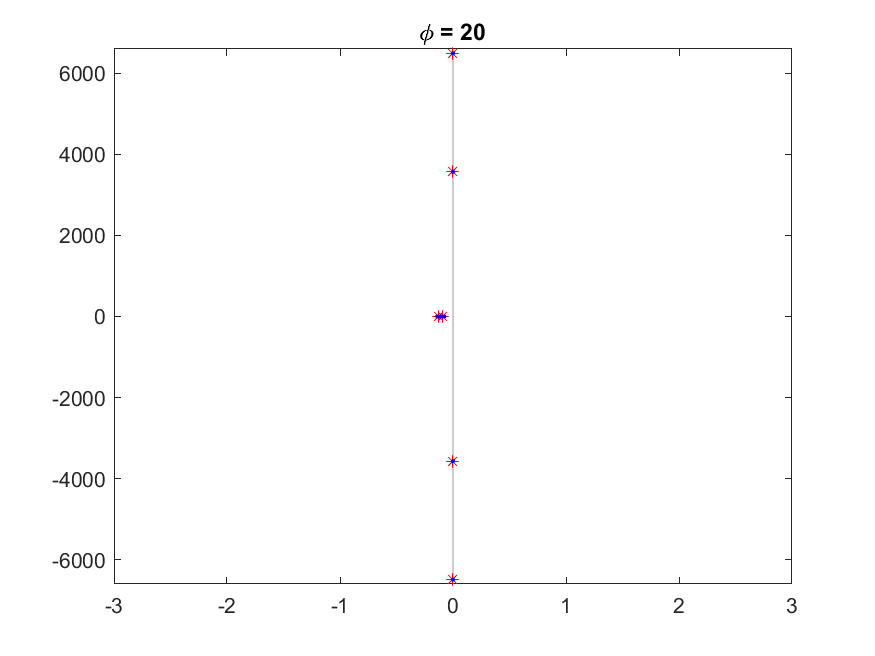}} 
\caption{Dissipative self-action: structured pseudospectrum of $A$ as $\phi$ varies.}
\label{fig: Autoazione dissipativa Pseudospettro di A al variare di N}
\end{figure}

Fixing $\chi=0.1 $ GPa and $\alpha=0$ GPa, as $\phi$ varies we obtain results collected in Table \ref{tab: Risult.Onde staz: autoazione dissipativa; VARIA N} and Figure \ref{fig: Autoazione dissipativa Pseudospettro di A al variare di N}.  Once again in figure \ref{fig: Autoazione dissipativa Pseudospettro di A al variare di N} we show the structured pseudospectra of $A$. Since $\chi<0.4$ GPa, instability is never reached. As $\phi$ grows, the imaginary part of the complex eigenvalues of $A$ diminishes. These eigenvalues correspond to damped oscillations and describe \emph{asymptotically stable states}: the origin is an attracting point in the phase space.
As $\varsigma$ grows, the damping decreases; also the real part of the eigenvalues grows as $\phi$ increases; in particular, when $\chi<0.4$ GPa, the spectral radius of $A$ goes to zero as $\phi$ increases. Correspondingly, we have non-oscillating solutions: the displacement tends exponentially to zero from the initial value, without oscillations.

\begin{table}
\caption{Dissipative self-action: spectrum of $A$ for some values of $\phi$.}
\label{tab: Risult.Onde staz: autoazione dissipativa; VARIA N}
\begin{center}
	\begin{tabular}{ |c|c|c|c| } 
		\hline
		\multicolumn{4}{|c|}{$\alpha=0$ GPa, $\chi=0.1 $ GPa} \\
		\hline
		$\phi$ & $18$ & $19$ & $20$ \\
		\hline
		\multirow{9}{4em}{$\sigma(A)$ $[\mathrm{rad/s}]$ } 
		& $-0.0004+6493\,i$&$-0.0001+6493\,i$&$-5\cdot 10^{-5}+6493\,i$\\
		& $-0.0004-6493\,i$&$-0.0001-6493\,i$&$-5\cdot 10^{-5}-6493\,i$\\
		& $-0.971$&$-0.357$&$-0.131$\\
		& $-0.001+3570\,i$&$-0.0004+3570\,i$&$-0.0002+3570\,i$\\
		& $-0.001-3570\,i$&$-0.0004-3570\,i$&$-0.0002-3570\,i$\\
		& $-0.001+3570\,i$&$-0.0004+3570\,i$&$-0.0002+3570\,i$\\
		& $-0.001-3570\,i$&$-0.0004-3570\,i$&$-0.0002-3570\,i$\\
		& $-0.668$&$-0.246$&$-0.09$\\
		& $-0.668$&$-0.246$&$-0.09$\\
		
		\hline
	\end{tabular}
\end{center}
\end{table}


\subsection{Simultaneous presence of conservative and dissipative components of $\mathbf{z}$}

\begin{figure}
\centering
\includegraphics[width=0.8\linewidth]{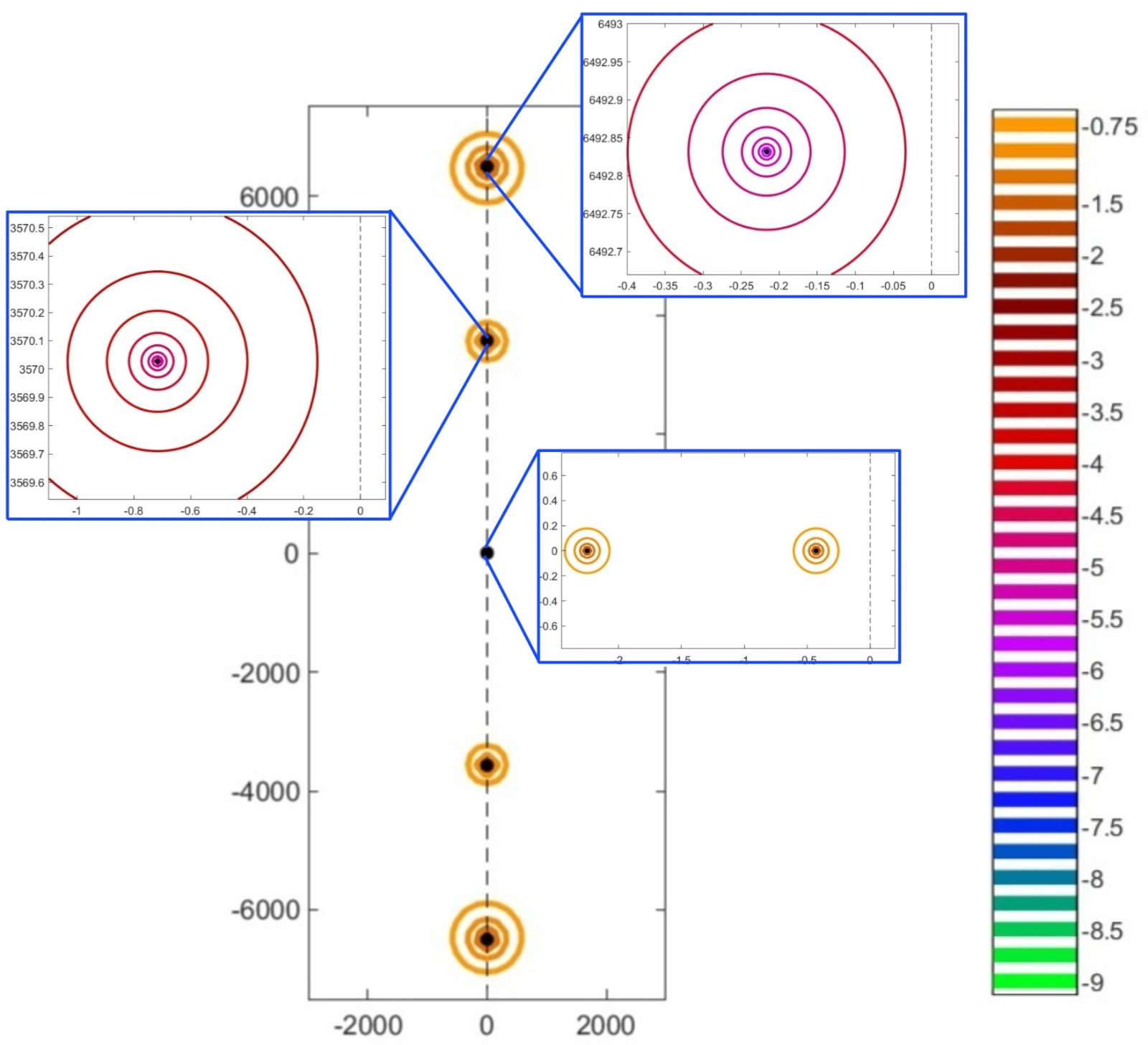}
\caption{ Complex $\varepsilon$-pseudospectrum of $A$ for $\chi=1.5$ GPa, $k_0=0.001$ GPa/m$^2$, $\phi=17$. The dashed line is the imaginary axis.}
\label{fig: Complex PS with zoom. Chi=1.5 GPa}
\end{figure}

When $\mathbf z = k_0  \nu + \varsigma \dot{\nu}$, we consider the matrix in formula \eqref{A-complete} with $\mathbb{K}_{i}$, $i=1,2,3$ given by formulas \eqref{K1-complete}, \eqref{K2-complete}, and \eqref{K3-complete}.
 It is genuinely non-normal, as confirmed by the measures of departure from normality: $\kappa_2(V) = 6.50\cdot 10^3$, $\dep^c(A)=1$ (up to six digits) and $\dep_F^H=4.58\cdot 10^7$. We have 
\begin{equation}
    1.52\cdot 10^7\leq {\dist}(A,\mathfrak{K}) \leq 4.48\cdot 10^7\;.
\end{equation}
Once again these measures remain almost constant for different values of $\chi$, $\varsigma$ and $k_0$. The departure from normality is the same as in the previous case. The spectrum of $A$ is stable under (generic) complex perturbations of the order of machine precision, as shown in figure \ref{fig: Complex PS with zoom. Chi=1.5 GPa}. Again, each color corresponds to the boundary of $\sigma_\varepsilon(A)$ for different values of $\varepsilon$ whose 10-based logarithm are displayed in the color-bar on the right.

Table \ref{tab: Risult.Onde staz: autoazione completa; varia chi k0=0.01} collects eigenvalues of $A$ as $\chi$ varies, by fixing $\alpha=0$ GPa, $k_0=0.01$ $\mathrm{GPa/m^2}$, and $\varsigma=0.178\,\mathrm{GPa\cdot s/m^2}$, while Table \ref{tab: Risult.Onde staz: autoazione completa; varia chi k0=0.1} shows values of $\sigma(A)$ for  $k_0=0.1$ $\mathrm{GPa/m^2}$, $\varsigma=0.178\,\mathrm{GPa\cdot s/m^2}$, and $\alpha=0$ GPa, as $\chi$ varies. Increasing the values of the conservative component of $\mathbf{z}$ has a \emph{stabilizing effect}. If  $\chi=3$ GPa, \emph{instability occurs} when $k_0=0.01$ $\mathrm{GPa/m^2}$, while \emph{stability is recovered} for $k_0=0.1$ $\mathrm{GPa/m^2}$.

\begin{table}
\caption{Spectrum of $A$ for some values of $\chi$, with $k_0=0.01$, $\mathrm{GPa/m^2}$, $\varsigma=0.178\,\mathrm{GPa\cdot s/m^2}$, and $\alpha=0$ GPa.}
\label{tab: Risult.Onde staz: autoazione completa; varia chi k0=0.01}
\begin{center}
	\begin{tabular}{ |c|c|c|c|c| } 
		\hline
		\multicolumn{5}{|c|}{$k_0=0.01$ $\mathrm{GPa/m^2}$, $\varsigma=0.178\,\mathrm{GPa\cdot s/m^2}$, $\alpha=0$ GPa.} \\
		\hline
		$\chi$ $[\mathrm{GPa}]$ & $0.5$ & $1$ & $1.7$ & $1.9$\\
		\hline
		\multirow{6}{2em}{$\sigma(A)$ $[\mathrm{rad/s}]$ } 
		& $-0.003+6492\,i$ & $-0.01+6492\,i$ & $-0.04+6492\,i$ & $-0.05+6492\,i$ \\
		& $-0.003-6492\,i$ & $-0.01-6492\,i$ & $-0.04-6492\,i$ & $-0.05-6492\,i$ \\
		& $-0.41$ & $-0.39$ & $-0.34$ & $-0.32$ \\
		& $-0.01+3570\,i$ & $-0.04+3570\,i$ & $-0.12+3570\,i$ & $-0.16+3570\,i$ \\
		& $-0.01-3570\,i$ & $-0.04-3570\,i$ & $-0.12-3570\,i$ & $-0.16-3570\,i$ \\
		& $-0.01+3570\,i$ & $-0.04+3570\,i$ & $-0.12+3570\,i$ & $-0.16+3570\,i$ \\
		& $-0.01-3570\,i$ & $-0.04-3570\,i$ & $-0.12-3570\,i$ & $-0.16-3570\,i$ \\
		& $-0.28$ & $-0.22$ & $-0.05$ & \textcolor{red}{+}$0.01$ \\
		& $-0.28$ & $-0.22$ & $-0.05$ & \textcolor{red}{+}$0.01$ \\
		\hline
	\end{tabular}
\end{center}
\end{table}
\begin{table}
\caption{Spectrum of $A$ for some values of $\chi$, with $k_0=0.1$, $\mathrm{GPa/m^2}$, $\varsigma=0.178\,\mathrm{GPa\cdot s/m^2}$, and $\alpha=0$ GPa.}
\label{tab: Risult.Onde staz: autoazione completa; varia chi k0=0.1}
\begin{center}
	\begin{tabular}{ |c|c|c|c|c| } 
		\hline
		\multicolumn{5}{|c|}{$k_0=0.1$ $\mathrm{GPa/m^2}$, $\varsigma=0.178\,\mathrm{GPa\cdot s/m^2}$, $\alpha=0$ GPa.} \\
		\hline
		$\chi$ $[\mathrm{GPa}]$ & $0.5$ & $1$ & $1.9$ & $3$\\
		\hline
		\multirow{6}{2em}{$\sigma(A)$ $[\mathrm{rad/s}]$ } 
		& $-0.003+6493\,i$ & $-0.013+6493\,i$ & $-0.047+6493\,i$ & $-0.117+6493\,i$ \\
		& $-0.003-6493\,i$ & $-0.013-6493\,i$ & $-0.047-6493\,i$ & $-0.117-6493\,i$ \\
		& $-0.911+0\,i$ & $-0.892+0\,i$ & $-0.824+0\,i$ & $-0.683+0\,i$ \\
		& $-0.011+3570\,i$ & $-0.043+3570\,i$ & $-0.156+3570\,i$ & $-0.388+3570\,i$ \\
		& $-0.011-3570\,i$ & $-0.043-3570\,i$ & $-0.156-3570\,i$ & $-0.388-3570\,i$ \\
		& $-0.011+3570\,i$ & $-0.043+3570\,i$ & $-0.156+3570\,i$ & $-0.388+3570\,i$ \\
		& $-0.011-3570\,i$ & $-0.043-3570\,i$ & $-0.156-3570\,i$ & $-0.388-3570\,i$ \\
		& $-0.785+0\,i$ & $-0.721+0\,i$ & $-0.496+0\,i$ & $-0.031+0\,i$ \\
		& $-0.785+0\,i$ & $-0.721+0\,i$ & $-0.496+0\,i$ & $-0.031+0\,i$ \\
		\hline
	\end{tabular}
\end{center}
\end{table}

\begin{figure}
\centering
\subfloat
{\includegraphics[width=0.49\textwidth]{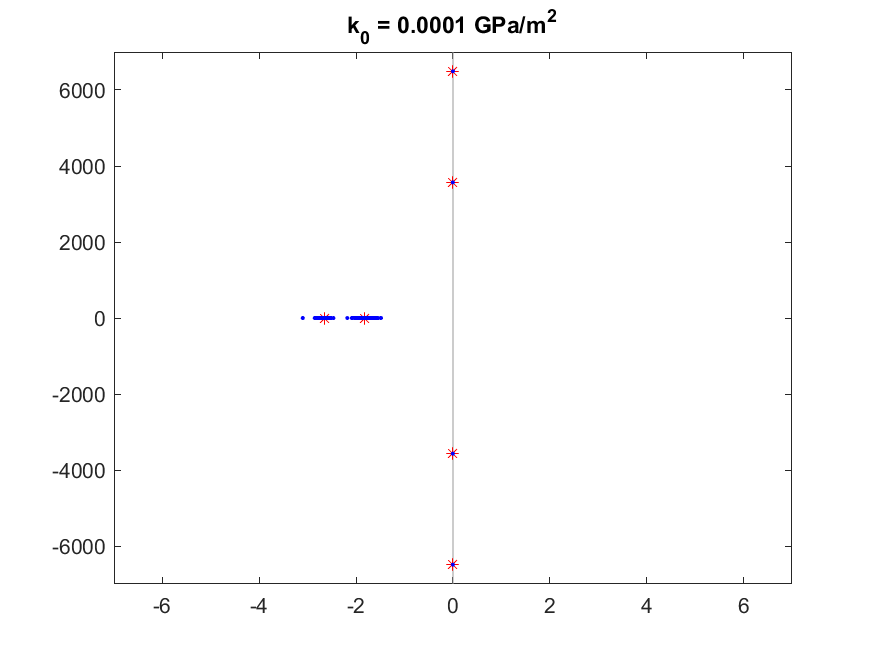}} \;
{\includegraphics[width=0.49\textwidth]{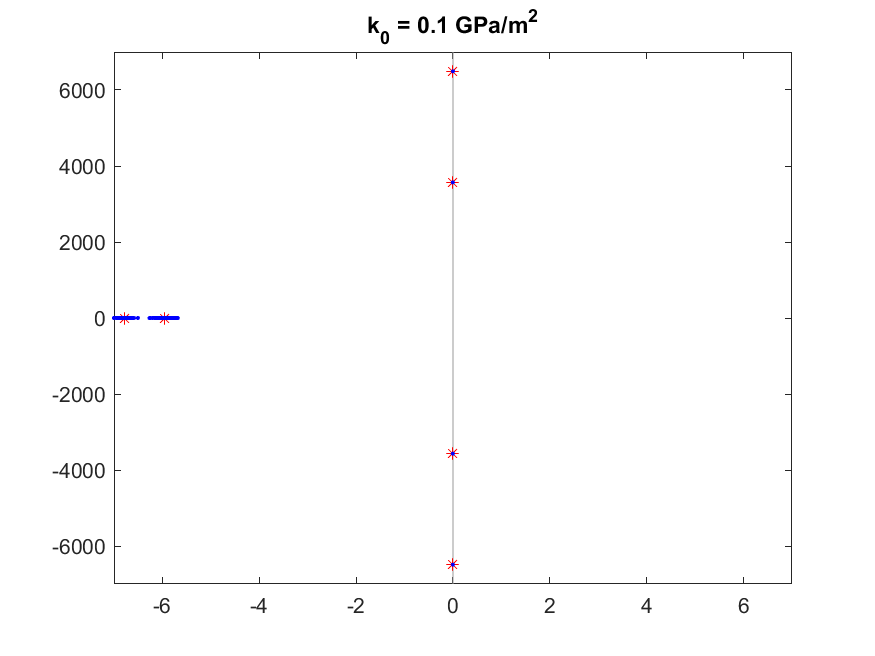}} 
\caption{Simultaneous presence of conservative and dissipative components of $\mathbf{z}$: structured pseudospectrum of $A$ as a function of $k_0$.}
\label{fig: Autoazione con componenti conservative e dissipative: pseudospettro di A al variare di k_0.}
\end{figure}

We fix $\chi=0.5$ GPa, $\alpha =0$ GPa, $\varsigma=0.178\,\mathrm{GPa\cdot s/m^2}$, and analyze the structured pseudospectrum of $A$ as $k_0$ varies. Results are in Figure \ref{fig: Autoazione con componenti conservative e dissipative: pseudospettro di A al variare di k_0.}: as $k_0$ grows, the three real eigenvalues of $A$ decrease; \emph{stability is assured}. 

\begin{figure}
\centering
\subfloat
{\includegraphics[width=0.49\textwidth]{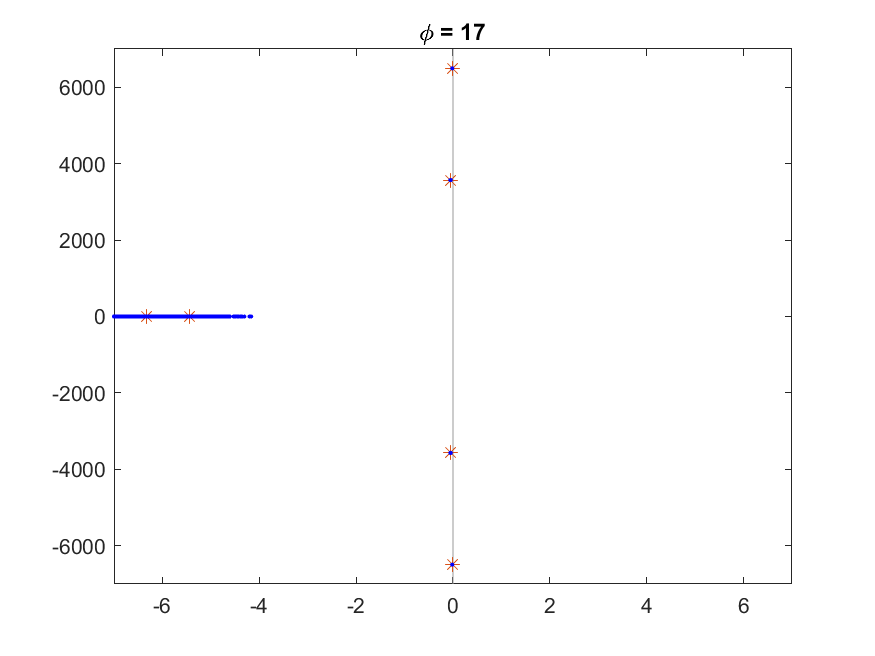}} \;
{\includegraphics[width=0.49\textwidth]{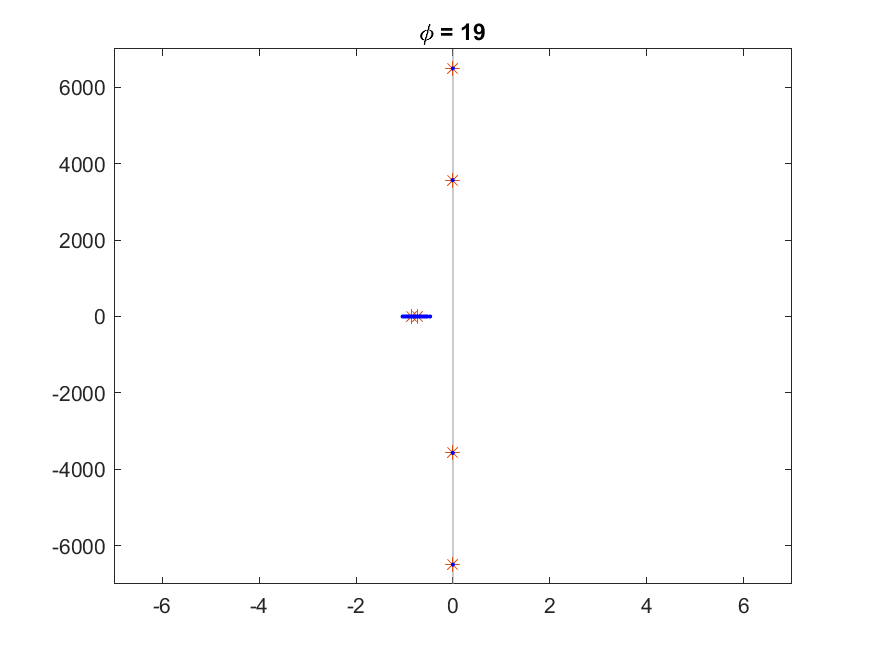}} 
\caption{Simultaneous presence of conservative and dissipative components of $\mathbf{z}$: structured pseudospectrum of $A$ as $\phi$ varies.}
\label{fig: Autoazione con componenti conservative e dissipative: pseudospettro di A al variare di N.}
\end{figure}

If we allow $\varsigma$ to vary, by changing only $\phi$, and fixing $\chi=0.4$ GPa, $\alpha=0$ GPa, $k_0=0.09$ $\mathrm{GPa/m^2}$, as $\varsigma$ grows the real eigenvalues of $A$ tend toward $0$ and, at the same time, the real part of the complex eigenvalues diminishes (see Figure \ref{fig: Autoazione con componenti conservative e dissipative: pseudospettro di A al variare di N.}). In this case, \emph{stability is assured} for every value of $\phi$ considered.

\begin{figure}
\centering
\subfloat
{\includegraphics[width=0.49\textwidth]{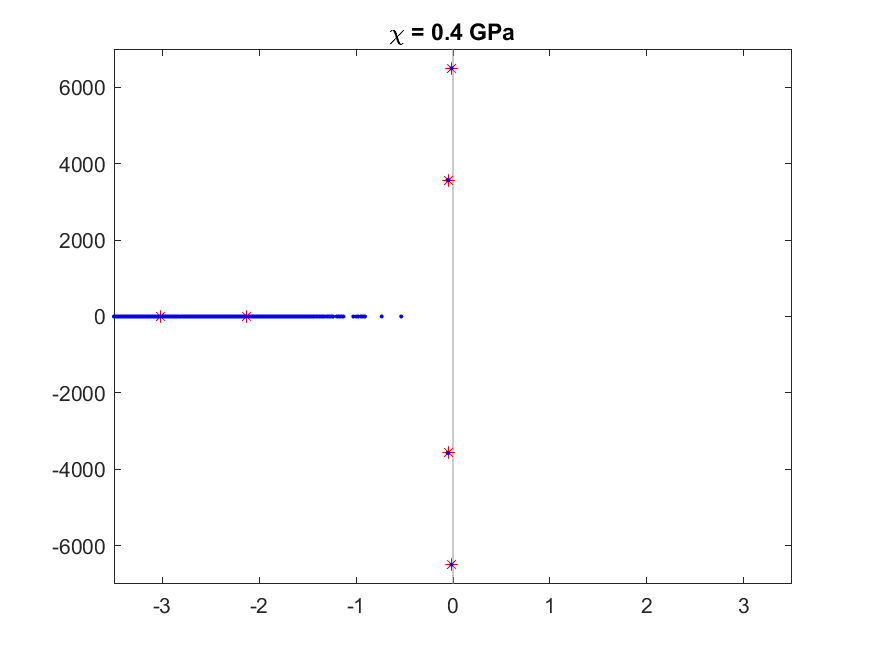}}\;
{\includegraphics[width=0.49\textwidth]{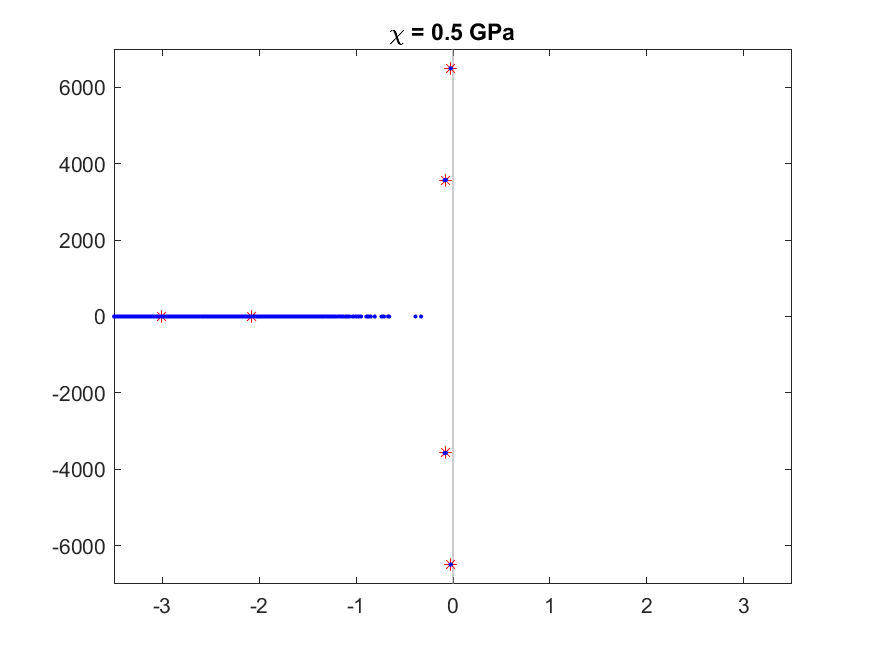}}\\
{\includegraphics[width=0.49\textwidth]{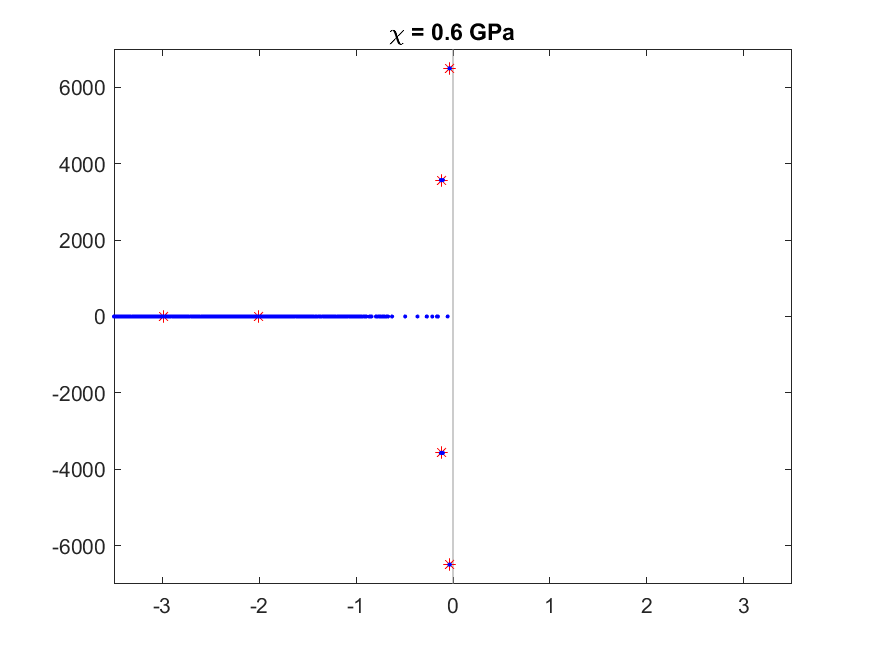}}\;
{\includegraphics[width=0.49\textwidth]{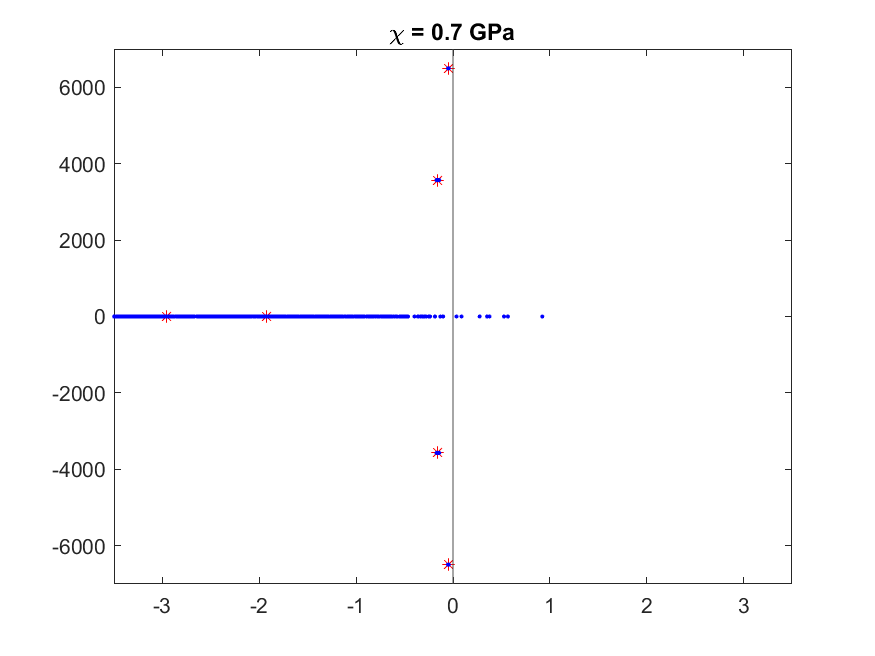}}\\
\caption{Simultaneous presence of conservative and dissipative components of $\mathbf{z}$: structured pseudospectrum of $A$ as $\chi$ varies, for $k_0=0.01$ GPa/m$^2$ and $\varsigma = 0.024$ GPa/s$\cdot$m$^2$ ($\phi=17$)}. Instability occurs for $\chi>0.6$ GPa.
\label{fig: Autoazione con componenti conservative e dissipative: pseudospettro di A al variare di chi. k0=0.01}
\end{figure}

\begin{figure}
\centering
\subfloat
{\includegraphics[width=0.49\textwidth]{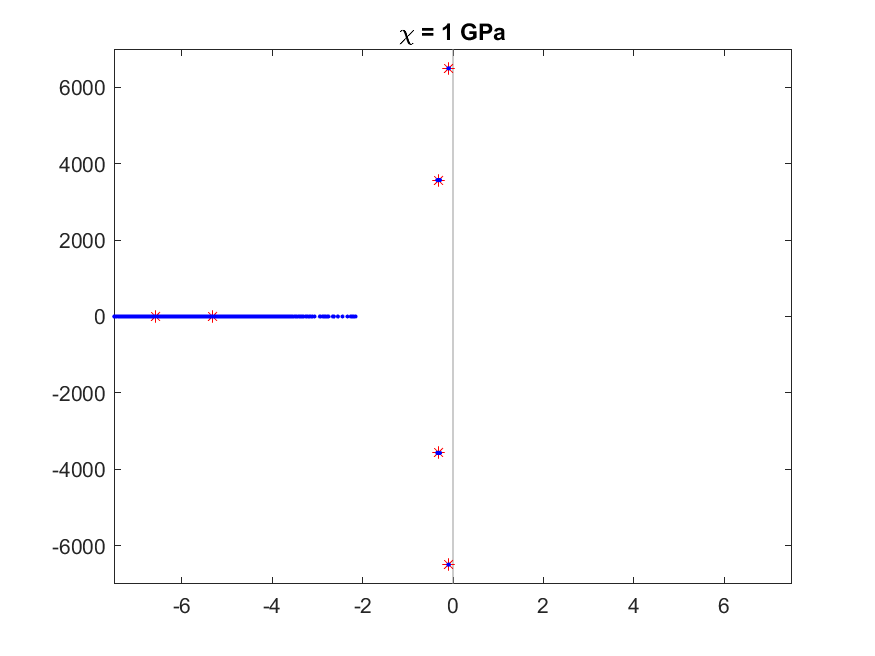}}\;
{\includegraphics[width=0.49\textwidth]{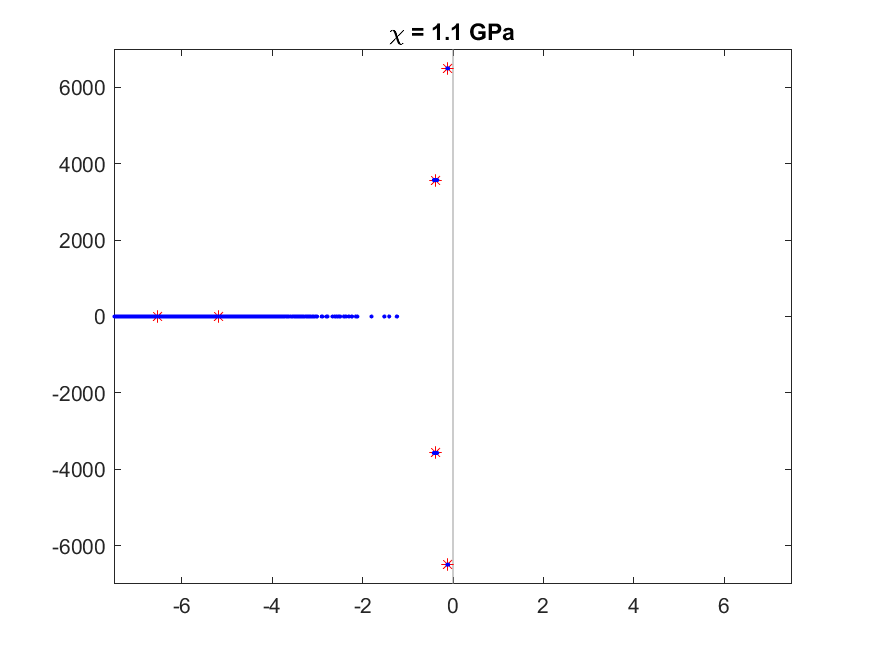}}\\
{\includegraphics[width=0.49\textwidth]{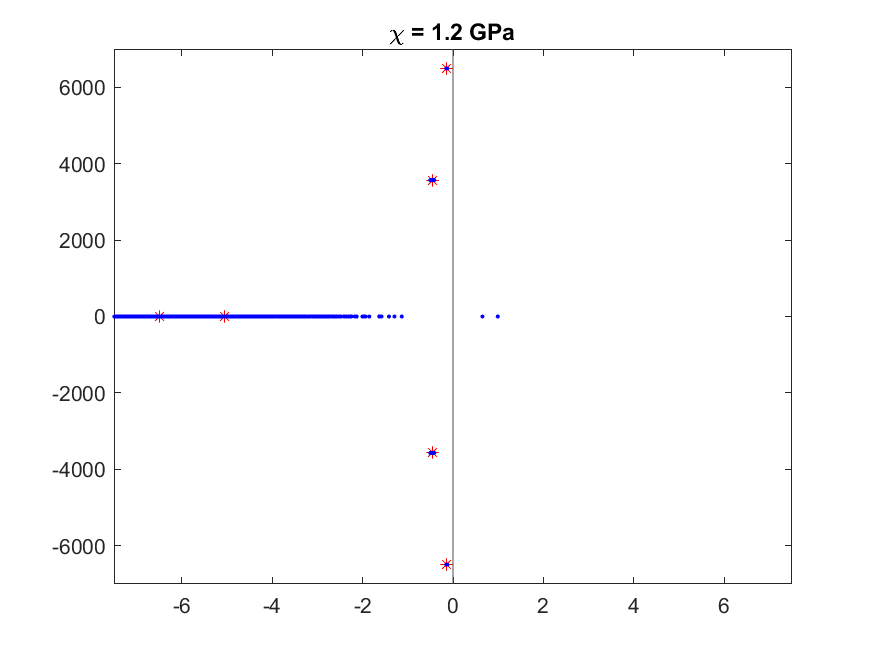}}\;
{\includegraphics[width=0.49\textwidth]{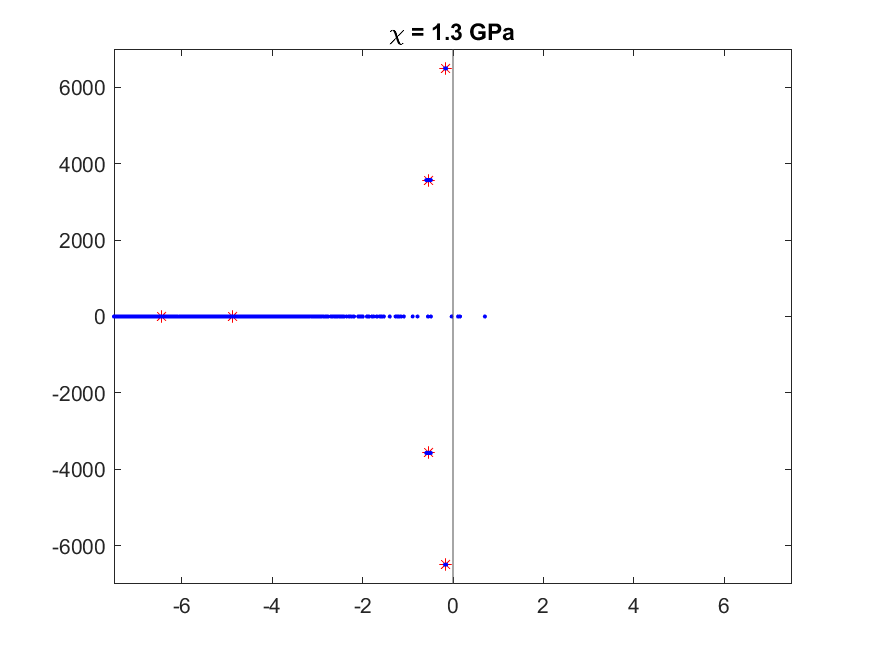}}\\
\caption{ Simultaneous presence of conservative and dissipative components of $\mathbf{z}$: structured pseudospectrum of $A$ as $\chi$ varies, for $k_0=0.1$ GPa/m$^2$ and $\varsigma = 0.024$ GPa/s$\cdot$m$^2$ ($\phi=17$).  Instability occurs for $\chi>1.1$ GPa. The conservative component of the phason self action stabilizes the system.}
\label{fig: Autoazione con componenti conservative e dissipative: pseudospettro di A al variare di chi. k0=0.1}
\end{figure}

Eventually, we allow $\chi$ to vary, keeping all the other parameters fixed. We set $\varsigma=0.178\,\mathrm{GPa\cdot s/m^2}$, $\alpha=0$ GPa, and $k_0=0.01$ $\mathrm{GPa/m^2}$. As $\chi$ grows, the three real eigenvalues of $A$ move towards the half-plane with positive real part and for $\chi>\sim 1.8$ GPa there is a \emph{Hopf-type transition towards instability}. By increasing $k_0$, \emph{instability occurs} for higher values of $\chi$; for example, for $k_0=0.1$ $\mathrm{GPa/m^2}$ \emph{instability occurs} when $\chi>3$ GPa, see table \ref{tab: Risult.Onde staz: autoazione completa; varia chi k0=0.1}. \\
However, the analysis of the structured pseudospectra shows that the real eigenvalues of $A$ are unstable under perturbations of the form \eqref{Perturbations E well scaled} for $\varepsilon$ corresponding to  $5\%$ of $\chi$.
The structured $\varepsilon$-pseudospectrum of $A$ crosses the imaginary axis for values of $\chi$ smaller than the ones predicted by the eigenvalues, as shown in figures \ref{fig: Autoazione con componenti conservative e dissipative: pseudospettro di A al variare di chi. k0=0.01} and \ref{fig: Autoazione con componenti conservative e dissipative: pseudospettro di A al variare di chi. k0=0.1} for two different values of $k_0$. Instability is reached already for $\chi \geq 0.7$ GPa when $k_0=0.01$ GPa/m$^2$ and for $\chi \geq 1.2$ GPa when $k_0=0.1$ GPa/m$^2$. We emphasize that increasing the conservative component of the phason self-action expands the admissible range of $\chi$.

\section{Concluding remarks}
When we analyze the dynamics of complex bodies, especially when microstructural diffusion is present, the evolution of waves with time-dependent amplitude may imply the emergence of non-normal matrices. The parameterized spectral analysis, which appears natural when some constitutive coefficients are experimentally unknown or are affected by uncertainties, 
may not be sufficient to clearly reveal stability and possible transition to unstable states. Exploring pseudospectra gives more reliable estimates, as those obtained in the present paper in the case of quasicrystals, for which constitutive parameters pertaining to the phason self-action are still unknown.
The same question emerges when we homogenize at continuum scale manufactured lattice-like structures at microscopic or mesoscopic scale (like in the case of printed metamaterials): a single lattice leads generically to a classical continuum scheme; multiple lattices imply often multi-field continuum schemes, those falling within the general model-building framework of the mechanics of complex materials (an offspring of this framework is what has been discussed here), where the occurrence of genuinely non-normal matrices in the material stability analysis cannot be excluded a priori. In all these cases, recourse to pseudospectra is necessary for a reliable evalutation of the stability of the material. Indeed, as shown in the previous analyses, there are circumstances in which pseudospectra predict instability when, instead, the spectral analysis indicates stability.

The recourse to pseudospectra requires the following steps:

\begin{itemize}
\item  First, generic (complex) pseudospectra should be examined, for a range of values of 
$\varepsilon$. If the eigenvalues are found to be stable under such perturbations, they will be so also for real and for structured perturbations involving only matrices which
preserve the  particular structure present (such as the block structure, the symmetry of the blocks, etc.).
\vskip6pt
\item If, on the other hand, the (generic) pseudospectra indicates potential instability of the spectrum, one should compute and analyze the structured pseudospectra.
\vskip6pt
\item In particular, the evaluation of the spectral pseudo-abscissa, namely the maximum real part of the points in the pseudospectra, is especially important from a robust stability point of view (see, e.g., \cite{GO}). Given a stable matrix, the minimal (with respect to the norm) perturbation that implies a pseudospectrum trespassing into the unstable region characterizes the type of stability pertaining to the original matrix.
\end{itemize}

The procedure is efficient and allows us to distinguish subtle behaviors that would otherwise be difficult to detect.

\FloatBarrier

\textbf{Acknowledgements}. We thank the three reviewers for their kind evaluations and the suggestions. This research falls within the activities of the Theoretical Mechanics Research Group of the Mathematics Research Center E.\ De Giorgi of the Scuola Normale Superiore in Pisa. GNFM-INDAM and GNCS-INDAM are acknowledged.

\end{document}